\input epsf
%
%
%
\def\unredoffs{} 

%
%
%
%
\newbox\leftpage \newdimen\fullhsize \newdimen\hstitle \newdimen\hsbody
\tolerance=1000\hfuzz=2pt
\catcode`\@=11 
%
\magnification=1200\unredoffs\baselineskip=16pt plus 2pt minus 1pt
\hsbody=\hsize \hstitle=\hsize 
%
%
%
\newcount\yearltd\yearltd=\year\advance\yearltd by -1900

%
%

\def\draftmode{\message{ DRAFTMODE }\def\draftdate{{\rm preliminary draft:
\number\month/\number\day/\number\yearltd\ \ \hourmin}}%
\headline={\hfil\draftdate}\writelabels\baselineskip=20pt plus 2pt minus 2pt
 {\count255=\time\divide\count255 by 60 \xdef\hourmin{\number\count255}
  \multiply\count255 by-60\advance\count255 by\time
  \xdef\hourmin{\hourmin:\ifnum\count255<10 0\fi\the\count255}}}
\def\nolabels{\def\wrlabeL##1{}\def\eqlabeL##1{}\def\reflabeL##1{}}
\def\writelabels{\def\wrlabeL##1{\leavevmode\vadjust{\rlap{\smash%
{\line{{\escapechar=` \hfill\rlap{\sevenrm\hskip.03in\string##1}}}}}}}%
\def\eqlabeL##1{{\escapechar-1\rlap{\sevenrm\hskip.05in\string##1}}}%
\def\reflabeL##1{\noexpand\llap{\noexpand\sevenrm\string\string\string##1}}}
\nolabels
%
\global\newcount\secno \global\secno=0
\global\newcount\meqno \global\meqno=1
\def\newsec#1{\global\advance\secno by1\message{(\the\secno. #1)}
\global\subsecno=0\eqnres@t\noindent{\bf\the\secno. #1}
\writetoca{{\secsym} {#1}}\par\nobreak\medskip\nobreak}
\def\eqnres@t{\xdef\secsym{\the\secno.}\global\meqno=1\bigbreak\bigskip}
\def\sequentialequations{\def\eqnres@t{\bigbreak}}\xdef\secsym{}
\global\newcount\subsecno \global\subsecno=0
\def\subsec#1{\global\advance\subsecno by1\message{(\secsym\the\subsecno. #1)}
\ifnum\lastpenalty>9000\else\bigbreak\fi
\noindent{\it\secsym\the\subsecno. #1}\writetoca{\string\quad
{\secsym\the\subsecno.} {#1}}\par\nobreak\medskip\nobreak}
\def\appendix#1#2{\global\meqno=1\global\subsecno=0\xdef\secsym{\hbox{#1.}}
\bigbreak\bigskip\noindent{\bf Appendix #1. #2}\message{(#1. #2)}
\writetoca{Appendix {#1.} {#2}}\par\nobreak\medskip\nobreak}
%
%
\def\eqnn#1{\xdef #1{(\secsym\the\meqno)}\writedef{#1\leftbracket#1}%
\global\advance\meqno by1\wrlabeL#1}
\def\eqna#1{\xdef #1##1{\hbox{$(\secsym\the\meqno##1)$}}
\writedef{#1\numbersign1\leftbracket#1{\numbersign1}}%
\global\advance\meqno by1\wrlabeL{#1$\{\}$}}
\def\eqn#1#2{\xdef #1{(\secsym\the\meqno)}\writedef{#1\leftbracket#1}%
\global\advance\meqno by1$$#2\eqno#1\eqlabeL#1$$}
%
\newskip\footskip\footskip14pt plus 1pt minus 1pt 
\def\footnotefont{\ninepoint}\def\f@t#1{\footnotefont #1\@foot}
\def\f@@t{\baselineskip\footskip\bgroup\footnotefont\aftergroup\@foot\let\next}
\setbox\strutbox=\hbox{\vrule height9.5pt depth4.5pt width0pt}
\global\newcount\ftno \global\ftno=0
\def\foot{\global\advance\ftno by1\footnote{$^{\the\ftno}$}}
%
\newwrite\ftfile
\def\footend{\def\foot{\global\advance\ftno by1\chardef\wfile=\ftfile
$^{\the\ftno}$\ifnum\ftno=1\immediate\openout\ftfile=foots.tmp\fi%
\immediate\write\ftfile{\noexpand\smallskip%
\noexpand\item{f\the\ftno:\ }\pctsign}\findarg}%
\def\footatend{\vfill\eject\immediate\closeout\ftfile{\parindent=20pt
\centerline{\bf Footnotes}\nobreak\bigskip\input foots.tmp }}}
\def\footatend{}
%
%
\global\newcount\refno \global\refno=1
\newwrite\rfile
\def\ref{[\the\refno]\nref}
\def\nref#1{\xdef#1{[\the\refno]}\writedef{#1\leftbracket#1}%
\ifnum\refno=1\immediate\openout\rfile=refs.tmp\fi
\global\advance\refno by1\chardef\wfile=\rfile\immediate
\write\rfile{\noexpand\item{#1\ }\reflabeL{#1\hskip.31in}\pctsign}\findarg}
\def\findarg#1#{\begingroup\obeylines\newlinechar=`\^^M\pass@rg}
{\obeylines\gdef\pass@rg#1{\writ@line\relax #1^^M\hbox{}^^M}%
\gdef\writ@line#1^^M{\expandafter\toks0\expandafter{\striprel@x #1}%
\edef\next{\the\toks0}\ifx\next\em@rk\let\next=\endgroup\else\ifx\next\empty%
\else\immediate\write\wfile{\the\toks0}\fi\let\next=\writ@line\fi\next\relax}}
\def\striprel@x#1{} \def\em@rk{\hbox{}}
\def\lref{\begingroup\obeylines\lr@f}
\def\lr@f#1#2{\gdef#1{\ref#1{#2}}\endgroup\unskip}
\def\semi{;\hfil\break}
\def\addref#1{\immediate\write\rfile{\noexpand\item{}#1}} 
\def\startrefs#1{\immediate\openout\rfile=refs.tmp\refno=#1}
\def\xref{\expandafter\xr@f}\def\xr@f[#1]{#1}
\def\refs#1{\count255=1[\r@fs #1{\hbox{}}]}
\def\r@fs#1{\ifx\und@fined#1\message{reflabel \string#1 is undefined.}%
\nref#1{need to supply reference \string#1.}\fi%
\vphantom{\hphantom{#1}}\edef\next{#1}\ifx\next\em@rk\def\next{}%
\else\ifx\next#1\ifodd\count255\relax\xref#1\count255=0\fi%
\else#1\count255=1\fi\let\next=\r@fs\fi\next}
%

%
\newwrite\ffile\global\newcount\figno \global\figno=1
\def\fig{fig.~\the\figno\nfig}
\def\nfig#1{\xdef#1{fig.~\the\figno}%
\writedef{#1\leftbracket fig.\noexpand~\the\figno}%
\ifnum\figno=1\immediate\openout\ffile=figs.tmp\fi\chardef\wfile=\ffile%
\immediate\write\ffile{\noexpand\medskip\noexpand\item{Fig.\ \the\figno. }
\reflabeL{#1\hskip.55in}\pctsign}\global\advance\figno by1\findarg}
\def\xfig{\expandafter\xf@g}\def\xf@g fig.\penalty\@M\ {}
\def\figs#1{figs.~\f@gs #1{\hbox{}}}
\def\f@gs#1{\edef\next{#1}\ifx\next\em@rk\def\next{}\else
\ifx\next#1\xfig #1\else#1\fi\let\next=\f@gs\fi\next}
\newwrite\lfile
{\escapechar-1\xdef\pctsign{\string\%}\xdef\leftbracket{\string\{}
\xdef\rightbracket{\string\}}\xdef\numbersign{\string\#}}

\def\writestop{\def\writestoppt{\immediate\write\lfile{\string\pageno%
\the\pageno\string\startrefs\leftbracket\the\refno\rightbracket%
\string\def\string\secsym\leftbracket\secsym\rightbracket%
\string\secno\the\secno\string\meqno\the\meqno}\immediate\closeout\lfile}}
\def\writestoppt{}\def\writedef#1{}
\def\seclab#1{\xdef #1{\the\secno}\writedef{#1\leftbracket#1}\wrlabeL{#1=#1}}
\def\subseclab#1{\xdef #1{\secsym\the\subsecno}%
\writedef{#1\leftbracket#1}\wrlabeL{#1=#1}}
\newwrite\tfile \def\writetoca#1{}
\def\leaderfill{\leaders\hbox to 1em{\hss.\hss}\hfill}
\def\writetoc{\immediate\openout\tfile=toc.tmp
   \def\writetoca##1{{\edef\next{\write\tfile{\noindent ##1
   \string\leaderfill {\noexpand\number\pageno} \par}}\next}}}

%
\catcode`\@=12 
%
\edef\tfontsize{\ifx\answ\bigans scaled\magstep3\else scaled\magstep4\fi}
\font\titlerm=cmr10 \tfontsize \font\titlerms=cmr7 \tfontsize
\font\titlermss=cmr5 \tfontsize \font\titlei=cmmi10 \tfontsize
\font\titleis=cmmi7 \tfontsize \font\titleiss=cmmi5 \tfontsize
\font\titlesy=cmsy10 \tfontsize \font\titlesys=cmsy7 \tfontsize
\font\titlesyss=cmsy5 \tfontsize \font\titleit=cmti10 \tfontsize
\skewchar\titlei='177 \skewchar\titleis='177 \skewchar\titleiss='177
\skewchar\titlesy='60 \skewchar\titlesys='60 \skewchar\titlesyss='60
\def\titlefont{\def\rm{\fam0\titlerm}
\textfont0=\titlerm \scriptfont0=\titlerms \scriptscriptfont0=\titlermss
\textfont1=\titlei \scriptfont1=\titleis \scriptscriptfont1=\titleiss
\textfont2=\titlesy \scriptfont2=\titlesys \scriptscriptfont2=\titlesyss
\textfont\itfam=\titleit \def\it{\fam\itfam\titleit}\rm}
 \ifx\answ\bigans\else scaled\magstep1\fi
\ifx\answ\bigans\else

 \font\absi=cmmi10 scaled\magstep1
\font\absis=cmmi7 scaled\magstep1 \font\absiss=cmmi5 scaled\magstep1
\font\abssy=cmsy10 scaled\magstep1 \font\abssys=cmsy7 scaled\magstep1
\font\abssyss=cmsy5 scaled\magstep1 
\skewchar\absi='177 \skewchar\absis='177 \skewchar\absiss='177
\skewchar\abssy='60 \skewchar\abssys='60 \skewchar\abssyss='60
\fi
\font\ninerm=cmr9 \font\sixrm=cmr6 \font\ninei=cmmi9 \font\sixi=cmmi6
\font\ninesy=cmsy9 \font\sixsy=cmsy6 \font\ninebf=cmbx9
\font\nineit=cmti9 \font\ninesl=cmsl9 \skewchar\ninei='177
\skewchar\sixi='177 \skewchar\ninesy='60 \skewchar\sixsy='60
\def\ninepoint{\def\rm{\fam0\ninerm}
\textfont0=\ninerm \scriptfont0=\sixrm \scriptscriptfont0=\fiverm
\textfont1=\ninei \scriptfont1=\sixi \scriptscriptfont1=\fivei
\textfont2=\ninesy \scriptfont2=\sixsy \scriptscriptfont2=\fivesy
\textfont\itfam=\ninei \def\it{\fam\itfam\nineit}\def\sl{\fam\slfam\ninesl}%
\textfont\bffam=\ninebf \def\bf{\fam\bffam\ninebf}\rm}
%
%
\def\noblackbox{\overfullrule=0pt}
\hyphenation{anom-aly anom-alies coun-ter-term coun-ter-terms}
\def\inv{^{\raise.15ex\hbox{${\scriptscriptstyle -}$}\kern-.05em 1}}

\def\Dsl{\,\raise.15ex\hbox{/}\mkern-13.5mu D} 
\def\dsl{\raise.15ex\hbox{/}\kern-.57em\partial}

\def\lspace{\ifx\answ\bigans{}\else\qquad\fi}
\def\lbspace{\ifx\answ\bigans{}\else\hskip-.2in\fi} 
\def\boxeqn#1{\vcenter{\vbox{\hrule\hbox{\vrule\kern3pt\vbox{\kern3pt
        \hbox{${\displaystyle #1}$}\kern3pt}\kern3pt\vrule}\hrule}}}
\def\mbox#1#2{\vcenter{\hrule \hbox{\vrule height#2in
                \kern#1in \vrule} \hrule}}  
%
   
 \def\CH{{\cal H}}

\def\darr#1{\raise1.5ex\hbox{$\leftrightarrow$}\mkern-16.5mu #1}
\def\chipp{\kappa}

\def\half{{\textstyle{1\over2}}} 
\def\roughly#1{\raise.3ex\hbox{$#1$\kern-.75em\lower1ex\hbox{$\sim$}}}
\hyphenation{Mar-ti-nel-li}

\def\1{\;1\!\!\!\! 1\;}

\def\ie{{\it i.e.}}

\def\etal{{\it et al.}}

\def\toinf#1{\mathrel{\mathop{\sim}\limits_{\scriptscriptstyle
{#1\rightarrow\infty }}}}

\def\frac#1#2{{{#1}\over {#2}}}
\def\half{\hbox{${1\over 2}$}}
\def\quarter{\hbox{${1\over 4}$}}
\def\smallfrac#1#2{\hbox{${{#1}\over {#2}}$}}

\catcode`@=11 
\def\slash#1{\mathord{\mathpalette\c@ncel#1}}
 \def\c@ncel#1#2{\ooalign{$\hfil#1\mkern1mu/\hfil$\crcr$#1#2$}}
\def\lsim{\mathrel{\mathpalette\@versim<}}
\def\gsim{\mathrel{\mathpalette\@versim>}}
 \def\@versim#1#2{\lower0.2ex\vbox{\baselineskip\z@skip\lineskip\z@skip
       \lineskiplimit\z@\ialign{$\m@th#1\hfil##$\crcr#2\crcr\sim\crcr}}}
\catcode`@=12 

\def\PR{{\it Phys.~Rev.~}}

\def\NP{{\it Nucl.~Phys.~}}

\def\PL{{\it Phys.~Lett.~}}

\def\SJNP{{\it Sov.~Jour.~Nucl.~Phys.~}}
\def\SPJETP{{\it Sov.~Phys.~J.E.T.P.~}}

\def\JHEP{{\it Jour.~High~Energy~Phys.~}}
\def\vol#1{{\bf #1}}\def\vyp#1#2#3{\vol{#1} (#2) #3}

\def\as{\alpha_s}

\def\Ai{\hbox{Ai}}
\def\ash{\widehat\alpha_s}

\noblackbox
\pageno=0\nopagenumbers\tolerance=10000\hfuzz=5pt
\baselineskip 12pt
\line{\hfill {\tt hep-ph/0306156}}
\line{\hfill CERN-TH/2003-019}
\line{\hfill Edinburgh 2003/08}
\line{\hfill IFUM-745/FT}
\vskip 12pt
\centerline{\titlefont An Anomalous Dimension for Small $x$ Evolution}
\vskip 36pt\centerline{Guido~Altarelli,$^{(a)}$
Richard D.~Ball$^{(b)}$ and Stefano Forte$^{(c)}$}
\vskip 12pt
\centerline{\it ${}^{(a)}$Theory Division, CERN}
\centerline{\it CH-1211 Gen\`eve 23, Switzerland}
\vskip 6pt
\centerline{\it ${}^{(b)}$School of Physics, University of Edinburgh}
\centerline{\it  Edinburgh EH9 3JZ, Scotland}
\vskip 6pt
\centerline {\it ${}^{(c)}$Dipartimento di  Fisica, Universit\`a di
Milano and}
\centerline{\it INFN, Sezione di Milano, Via Celoria 16, I-20133 Milan, Italy}
\vskip 50pt
\centerline{\bf Abstract}
{\narrower\baselineskip 10pt
\medskip\noindent 
We construct an anomalous dimension for small $x$ evolution which goes
beyond standard fixed order perturbative evolution by including
resummed small $x$ logarithms deduced from the leading order BFKL
equation with running coupling. Surprisingly, we find that once 
running coupling effects are properly taken into account, the 
leading approximation is very close to standard perturbative 
evolution in the range of $x$ accessible at HERA, in overall agreement 
with the data, with no need for phenomenological parameters to 
summarise subleading effects. We also show that further corrections 
due to subleading small $x$ logarithms derived from the Fadin-Lipatov 
kernel can be kept under control, but that they involve substantial 
resummation ambiguities which limit their practical usefulness. 
}
\vfill
\line{CERN-TH/2003-019\hfill }
\line{June 2003\hfill}
\eject \footline={\hss\tenrm\folio\hss}

\lref\dasnew{
S.~Forte and R.~D.~Ball,
{\tt hep-ph/0109235.}
}
\lref\haidt{D.~Haidt, Talk at DIS 2003, St.~Petersburg, Russia, April
2003.}
\lref\glap{
V.N.~Gribov and L.N.~Lipatov,
\SJNP\vyp{15}{1972}{438}\semi  
L.N.~Lipatov, \SJNP\vyp{20}{1975}{95}\semi    
G.~Altarelli and G.~Parisi,
\NP\vyp{B126}{1977}{298}\semi  
see also
Y.L.~Dokshitzer,
{\it Sov.~Phys.~JETP~}\vyp{46}{1977}{691}.} 
\lref\nlo{G.~Curci, W.~Furma\'nski and R.~Petronzio,
\NP\vyp{B175}{1980}{27}\semi 
E.G.~Floratos, C.~Kounnas and R.~Lacaze,
\NP\vyp{B192}{1981}{417}.} 
\lref\nnlo{S.A.~Larin, T.~van~Ritbergen, J.A.M.~Vermaseren,
\NP\vyp{B427}{1994}{41}\semi  
S.A.~Larin \etal, \NP\vyp{B492}{1997}{338}.} 
\lref\bfkl{L.N.~Lipatov,
\SJNP\vyp{23}{1976}{338}\semi 
 V.S.~Fadin, E.A.~Kuraev and L.N.~Lipatov,
\PL\vyp{60B}{1975}{50}; 
 {\it Sov. Phys. JETP~}\vyp{44}{1976}{443}; 
\vyp{45}{1977}{199}\semi 
 Y.Y.~Balitski and L.N.Lipatov,
\SJNP\vyp{28}{1978}{822}.} 
\lref\CH{
S.~Catani and F.~Hautmann,
\PL\vyp{B315}{1993}{157}; 
\NP\vyp{B427}{1994}{475}.} 
\lref\fl{V.S.~Fadin and L.N.~Lipatov,
\PL\vyp{B429}{1998}{127}\semi  
V.S.~Fadin et al, \PL\vyp{B359}{1995}{181}; 
\PL\vyp{B387}{1996}{593}; 
\NP\vyp{B406}{1993}{259}; 
\PR\vyp{D50}{1994}{5893}; 
\PL\vyp{B389}{1996}{737};  
\NP\vyp{B477}{1996}{767};  
\PL\vyp{B415}{1997}{97};  
\PL\vyp{B422}{1998}{287}.} 
\lref\cc{G.~Camici and M.~Ciafaloni,
\PL\vyp{B412}{1997}{396}; 
\PL\vyp{B430}{1998}{349}.} 
\lref\dd{V.~del~Duca, \PR\vyp{D54}{1996}{989};
\PR\vyp{D54}{1996}{4474}\semi 
V.~del~Duca and C.R.~Schmidt,
\PR\vyp{D57}{1998}{4069}\semi 
Z.~Bern, V.~del~Duca and C.R.~Schmidt,
\PL\vyp{B445}{1998}{168}.}
\lref\ross{
D.~A.~Ross,
Phys.\ Lett.\ B {\bf 431}, 161 (1998) 
}
\lref\jar{T.~Jaroszewicz,
\PL\vyp{B116}{1982}{291}.}
\lref\ktfac{S.~Catani, F.~Fiorani and G.~Marchesini,
\PL\vyp{B336}{1990}{18}\semi 
S.~Catani et al.,
\NP\vyp{B361}{1991}{645}.}
\lref\summ{R.~D.~Ball and S.~Forte,
\PL\vyp{B351}{1995}{313}\semi  
R.K.~Ellis, F.~Hautmann and B.R.~Webber,
\PL\vyp{B348}{1995}{582}.}
\lref\afp{R.~D.~Ball and S.~Forte,
\PL\vyp{B405}{1997}{317}.}
\lref\DGPTWZ{A.~De~R\'ujula {\it et al.},
\PR\vyp{D10}{1974}{1649}.}
\lref\das{R.~D.~Ball and S.~Forte,
\PL\vyp{B335}{1994}{77}; 
\vyp{B336}{1994}{77}\semi 
{\it Acta~Phys.~Polon.~}\vyp{B26}{1995}{2097}.}
\lref\kis{
See {\it e.g.}  R.~K.~Ellis, W.~J.~Stirling and B.~R.~Webber,
``QCD and Collider Physics'' (C.U.P., Cambridge 1996).}
\lref\hone{H1 Collab., {\it Eur.\ Phys.\ J.} C {\bf 21} (2001)
33.}
\lref\zeus{ZEUS Collab., {\it Eur.\ Phys.\ J.}
 C {\bf 21} (2001) 443.} 
\lref\mom{R.~D.~Ball and S.~Forte, {\it Phys. Lett.} {\bf
B359}, 362 (1995).}
\lref\bfklfits{R.~D.~Ball and S.~Forte,
{\tt hep-ph/9607291}\semi 
I.~Bojak and M.~Ernst, \PL\vyp{B397}{1997}{296};
\NP\vyp{B508}{1997}{731}\semi
J.~Bl\"umlein  and A.~Vogt,
\PR\vyp{D58}{1998}{014020}.}
\lref\flph{R.~D.~Ball  and S.~Forte,
{\tt hep-ph/9805315}\semi 
J. Bl\"umlein et al.,
{\tt hep-ph/9806368}.}
\lref\salam{G.~Salam, \JHEP\vyp{9807}{1998}{19}.}
\lref\sxap{R.~D.~Ball and S.~Forte,
\PL\vyp{B465}{1999}{271}.}
\lref\sxres{G. Altarelli, R.~D. Ball and S. Forte,
\NP{\bf B575}, 313 (2000);  
see also {\tt hep-ph/0001157}.
}
\lref\sxphen{G. Altarelli, R.~D.~Ball and S. Forte,
\NP\vyp{B599}{2001}{383};  
see also {\tt hep-ph/0104246}.}  
\lref\ciaf{M.~Ciafaloni and D.~Colferai,
\PL\vyp{B452}{1999}{372}\semi 
M.~Ciafaloni, G.~Salam and  D.~Colferai,
{\tt hep-ph/9905566}.}  
\lref\Liprun{L.N.~Lipatov,
\SPJETP\vyp{63}{1986}{5}.}
\lref\ColKwie{
J.~C.~Collins and J.~Kwiecinski, \NP\vyp{B316}{1989}{307}.}
\lref\CiaMue{
M.~Ciafaloni, M.~Taiuti and A.~H.~Mueller,
{\tt hep-ph/0107009}.
}
\lref\ciafac{
M.~Ciafaloni, D.~Colferai and G.~P.~Salam,
JHEP {\bf 0007}  (2000) 054
}
\lref\ciafrun{G.~Camici and M.~Ciafaloni,
\NP\vyp{B496}{1997}{305}.}
\lref\Haak{L.~P.~A.~Haakman, O.~V.~Kancheli and
J.~H.~Koch \NP\vyp{B518}{1998}{275}.} 
\lref\Bartels{N. Armesto, J. Bartels and M.~A.~Braun,
\PL\vyp{B442}{1998}{459}.} 
\lref\Thorne{R.~S.~Thorne,
\PL\vyp{B474}{2000}{372}; {\it Phys.\ Rev.} {\bf D64} (2001) 074005.
}
\lref\anders{
J.~R.~Andersen and A.~Sabio Vera,
{\tt arXiv:hep-ph/0305236.}
}
\lref\mf{
S.~Forte and R.~D.~Ball,
AIP Conf.\ Proc.\  {\bf 602} (2001) 60
{\tt hep-ph/0109235.}
}
\lref\sxrun{
G.~Altarelli, R.~D.~Ball and S.~Forte,
Nucl.\ Phys.\ B {\bf 621} (2002)  359.
}
\lref\newciaf{M.~Ciafaloni et al., {\tt hep-ph/0305254}.
}
\lref\ciafrc{M.~Ciafaloni, M.~Taiuti and A.~H.~Mueller,
{\it Nucl.\ Phys.}  {\bf B616} (2001) 349\semi 
M.~Ciafaloni et al., {\it Phys. Rev.} {\bf D66} (2002)
054014 
}
\newsec{Introduction}
\noindent In recent years the theory of scaling violations for deep inelastic
structure functions at small $x$ has
attracted considerable interest, prompted by the
experimental information coming from
HERA~\refs{\hone,\zeus}. New effects beyond the low--order perturbative
approximation~\refs{\glap\nlo{--}\nnlo} to
anomalous dimensions  or splitting functions should become important at
small-$x$. The BFKL
approach~\refs{\bfkl\CH\fl\cc{--}\dd}  provides in principle a tool for the
determination of the small-$x$ improvements
of the anomalous dimensions~\refs{\jar,\ktfac}. However, no major deviation of
the data from a standard
next--to--leading order perturbative treatment of  scaling
violations~\refs{\DGPTWZ,\das,\hone,\zeus} has been found, and
a straightforward~\summ\ inclusion of  small-$x$ logarithms is in fact ruled
out by the
data~\refs{\bfklfits,\flph}. By now the origin of this situation has been
mostly
understood~\refs{\salam\sxres\sxphen{--}\ciaf}.

The BFKL kernel $\chi(\alpha_s,M)$, as is well known, has been computed to
next-to-leading accuracy (NLO):
\eqn\chidef{
\chi(M,\alpha_s)=\alpha_s \chi_0(M)~+~\alpha_s^2 \chi_1(M)~+~\dots . } The
problem is how to use the information contained
in
$\chi_0$ and $\chi_1$ in order to improve the splitting function derived from
the perturbative leading singlet anomalous
dimension function $\gamma(\alpha_s,N)$ which is known~\nlo\ up to
NLO in $\alpha_s$:
\eqn\gammadef{
\gamma(N,\alpha_s)=\alpha_s \gamma_0(N)~+~\alpha_s^2
\gamma_1(N)~+~\dots ,} in such a way that the improved splitting function
remains a good approximation down to small values
of $x$. This can be accomplished by exploiting the
fact~\refs{\afp\sxres\sxphen} that the solutions of the BFKL and GLAP
equations coincide at leading twist if their respective evolution kernels are
related by a ``duality'' relation.  In the fixed coupling limit
the duality relation is simply given by:
\eqn\dualdef{
\chi(\gamma(N,\as),\as)=N.} The splitting function then will contain
 all relative corrections of order
$(\alpha_s \log{1/x})^n$, derived from
$\chi_0(M)$, and of order
$\alpha_s(\alpha_s \log{1/x})^n$, derived from $\chi_1(M)$.

The early wisdom on how to implement the information from $\chi_0$ was
completely shaken by the much
softer behaviour of the data at small $x$ and by the computation of
$\chi_1$~\refs{\fl\cc{--}\dd}, which showed that the naive expansion for the
improved anomalous dimension had a hopelessly
bad behaviour. In refs.~\refs{\sxphen,\sxres} we have shown that, as a
consequence of the physical requirement of momentum 
conservation,  
these problems
are cured if the small-$x$ resummation is
combined with the standard resummation of collinear singularities, by
constructing a `double-leading' perturbative
expansion. However, the next-to-leading correction
remains large, and it qualitatively changes the asymptotic small-$x$ behaviour
of structure functions by changing it from
$x^{-\lambda_0}$ to $x^{-\lambda}~=~x^{-\lambda_0} e^{\Delta \lambda
\xi}~\approx ~x^{-\lambda_0}[1+\Delta \lambda \xi+....]$, with:
\eqn\lambdadef{\lambda=\lambda_0+\Delta
\lambda,~~~~~\lambda_0=\as\chi_0(\half)=\as
c_0,~~~~~\Delta\lambda=\as^2\chi_1(\half)+\cdots. } The effects of this
correction can be resummed, but there are ambiguities in the procedure which
entail the dependence of the results on free
parameters, or, alternatively, model assumptions.

A traditionally delicate point of the small-$x$ resummation is the treatment of
the running of the
coupling~\refs{\ciafrun\Haak\Bartels\Thorne\ciafrc{--}\anders}. While it has been known for
some time~\refs{\sxap\sxres\sxphen} that
running coupling effects can be included perturbatively order by order at
small $x$, in a recent paper~\sxrun\  we have
shown that an all--order resummation of running coupling effects at
small $x$
is desirable and can in fact be
accomplished. Indeed, in a perturbative approach, running coupling terms can be
included by adding effective subleading
$\Delta \chi_i$ contributions to the BFKL kernels $\chi_1, \chi_2, \dots$,
which turn out to have singularities at $M=1/2$.
These singularities correspond~\refs{\sxap,\sxrun} to an enhancement by powers
of $\ln 1/x$ of the associated splitting
functions, which may offset the perturbative suppression by powers of $\as$. In
ref.~\sxrun\ we have shown that the
all-order resummation of these contributions is  fully compatible (up to higher
twist corrections) with the standard
factorized perturbative evolution of parton distributions, and can thus be
performed at the level of splitting functions.
The resummed splitting functions remain smooth in the small-$x$ limit, despite
the
well-known~\refs{\ciafrun\Haak\Bartels{--}\Thorne} fact that  the naive running
coupling BFKL solution instead displays
oscillatory instabilities at small $x$.

The purpose of this paper is to discuss the physical impact of this running
coupling resummation, and describe in detail
its phenomenological implementation. We will argue that we now know the way the
information
contained in
$\chi_0(M)$ should be used in order to construct a better first approximation
for the improved anomalous dimension.
Indeed, we find that, once running coupling effects are properly included in
the improved anomalous dimension, the
asymptotic behavior near $x=0$ is much softened with respect to the naive
Lipatov exponent. Hence, the corresponding
dramatic rise of structure functions at small $x$, which is phenomenologically
ruled out, is replaced by a milder rise,
whose steepness is determined by the Lipatov exponent. It then follows that
a leading--order approximation based on the standard BFKL kernel
$\chi_0$ is phenomenologically viable. 

Our proposed leading approximation is of
a reasonably simple explicit form, directly
suitable for the determination of scaling violations and fits to
the data, contains no
free parameters, and is in
agreement with the general trend of the data. We find it remarkable that such a
first
approximation indeed exists and apparently works so well, so that we devote a
good fraction of this article to describing it
and its physical foundation. We then construct and discuss a perturbative
expansion based upon it, and show in particular
that the next order correction, which includes the effects of
the subleading kernel
$\chi_1$,  leads to small corrections for a reasonable range of the
resummation parameters. However, 
some parameter dependence and resummation
ambiguities are necessarily present at this level, and in fact at
next-to-leading order the
associated ambiguity is of the same size as the correction itself. 
The fact that the corrections are
small but ambiguous justifies the
use of the simple leading--order
approximation for practical purposes.

\newsec{Notation and Basic Formalism}
\noindent

The behaviour of structure functions at small $x$ is dominated by the large
eigenvalue of evolution in the  singlet sector.
Thus we consider the singlet parton density
\eqn\Gdef{  G(\xi,t)=x[g(x,Q^2)+k_q\otimes q(x,Q^2)], }  where $\xi=\log{1/x}$,
$t=\log{Q^2/\mu^2}$,
$g(x,Q^2)$ and $q(x,Q^2)$ are the gluon and singlet quark parton  densities,
respectively, and $k_q$ is such that, for each
moment
\eqn\Nmom{  G(N,t)=\int^{\infty}_{0}\! d\xi\, e^{-N\xi} G(\xi,t), }  the
associated anomalous dimension
$\gamma(\as(t),N)$ corresponds  to the largest eigenvalue in the singlet
sector. The generalization to the full
two--by--two matrix of anomalous dimensions is discussed in detail in
ref.~\sxphen.

At large $t$ and fixed $\xi$ the evolution equation  in $N$-moment space is
then
\eqn\tevol{
\frac{d}{dt}G(N,t)=\gamma(\as(t),N) G(N,t), }  where $\as(t)$ is the running
coupling. The anomalous dimension  is
completely known   at one-- and two--loop level, as given in eq.~\gammadef. The
corresponding splitting function is
related by a Mellin  transform to $\gamma(\as,N)$:
\eqn\psasy{
\gamma(\as,N) =\int^1_0\!dx\,x^N\!P(\as,x). }  
Small $x$ for the splitting
function corresponds to small $N$ for  the
anomalous dimension: more precisely
$P\sim 1/x(\log(1/x))^n$ corresponds to $\gamma\sim n!/N^{n+1}$.  Even assuming
that a leading twist description of scaling
violations is still valid in  some range of small $x$, as soon as
$x$ is small enough that
$\as \xi\sim 1$, with $\xi=\log{1/x}$, all terms of order
$(\as/x) (\as \xi)^n$ (LLx) and $\as(\as/x)  (\as \xi)^n$ (NLLx)  which are
present in the splitting functions must be
considered in order to achieve an accuracy up  to terms of order
$\as^2(\as/x) (\as \xi)^n$ (NNLLx).

As is well known, these terms can be derived from the knowledge of the kernel
$\chi(\as,M)$, given in NLO approximation in eq.~\chidef, of the BFKL
$\xi$--evolution equation
\eqn\xevol{
\frac{d}{d\xi}G(\xi,M)=\chi(\as,M) G(\xi,M), }  which is satisfied at large
$\xi$ by the inverse Mellin  transform of the parton distribution
\eqn\Mmom{  G(\xi,M)=\int^{\infty}_{-\infty}\! dt\, e^{-Mt} G(\xi,t). }  This
derivation was originally performed~\jar\ at
LLx by assuming the common validity of eq.~\xevol\ and eq.~\tevol\ in the
region where $Q^2$ and $\xi$ are both large.
However, it was more recently realized~\refs{\afp,\sxres,\sxphen}
 that the solution of eq.~\xevol\ coincides generally with that of eq.~\tevol,
up to higher twist corrections, provided
only that the kernel of the former is related to that of the latter  by a
`duality' relation, and boundary conditions are
suitably matched. This implies that  the domains of validity of these two
equations are in fact the same in perturbation
theory,  up to power--suppressed corrections.

The derivation of duality is simplest when the coupling does not run, in which
case the relation between the kernels of the
two equations is given by eq.~\dualdef.  A simple proof of this statement was
given in ref.~\sxres.  Expanding
$\gamma(\as,N)$ in powers of $\as$ at fixed
$\as/N$
\eqn\sxexp {\gamma(\as,N)=\gamma_s(\as/N)+\as\gamma_{ss}(\as/N)+\dots,} and
$\chi(\as,M)$ in powers of
$\as$ at fixed $M$
\eqn\chiexp{\chi(\as,M)=\as\chi_0(M)+\as^2 \chi_1(M)+\dots,}  we then find that
$\chi_0$ determines $\gamma_s(\as/N)$, while $\as \chi_1$ leads to
$\as\gamma_{ss}(\as/N)$:\eqnn\lodual\eqnn\nlodual
$$\eqalignno{
\chi_{0}(\gamma_{s}(\smallfrac{\as}{N}))&={N\over\as},&\lodual\cr
\gamma_{ss}(\smallfrac{\as}{N})&=
-\frac{\chi_{1}(\gamma_{s}(\smallfrac{\as}{N}))}
{\chi'_{0}(\gamma_{s}(\smallfrac{\as}{N}))}.&\nlodual\cr}$$ Upon Mellin
inversion, $\gamma_s(\as/N)$ and $\as
\gamma_{ss}(\as/N)$  correspond respectively to all terms of order $(\as/x)
(\as \xi)^n$ and
$\as(\as/x)(\as \xi)^n$
 in the splitting functions.

When one goes beyond LLx, i.e. beyond the leading--order approximation for
$\chi$, the  running of the  coupling cannot be neglected, and the duality
relation must be re--examined. Indeed,  in $M$
space the usual running coupling $\as(t)$ becomes a differential operator and,
by taking only the one-loop beta function
into account, one has:
\eqn\ashdef{
\ash = \frac{\as}{1-\beta_0 \as \smallfrac{d}{dM}}+\cdots, } where $\beta_0$ is
the first coefficient of the
$\beta$-function (so
$\beta=-\beta_0\as^2+\cdots$ and $\beta_0=0.663146$... for 4 flavours),  with
the obvious generalization to higher orders.
Hence, assuming the coupling to run in the usual way with $Q^2$, the
$\xi$-evolution equation eq.~\xevol\  becomes~\Liprun
\eqn\xevolrun{
\frac{d}{d\xi}G(\xi,M)=\chi(\ash,M) G(\xi,M), }  where the derivative with
respect to $M$ acts on everything to the right,
and
$\chi$ may be expanded as in eq.~\chidef\ keeping the powers  of $\ash$ on the
left. In ref.~\sxrun\ we have shown that eq.~\xevolrun\
can indeed be viewed as an alternative representation of the standard
renormalization--group equation.

It is clear from eq.~\xevolrun\ that  running coupling effects begin at NLLx.
To NLLx one finds~\sxap\ that the solution of eq.~\xevolrun\
is again the same as that of a dual $t$--evolution equation~\tevol, provided
the fixed--coupling duality relation
eq.~\dualdef\ is modified by letting $\as\to\as(t)$, and then by adding to
$\gamma_{ss}$ eq.~\nlodual\ an extra term~\ciafrun\
$\Delta \gamma_{ss}$  proportional to $\beta_0$:
\eqn\deltag{
\Delta \gamma_{ss}(\smallfrac{\as}{N})=-\beta_0
\frac{\chi_0''(\gamma_{s})\chi_0(\gamma_{s})} {2\chi_0'^2(\gamma_{s})}.    }
Equivalently,  the duality relation
eq.~\dualdef\ can be formally preserved, provided that $\as\to\as(t)$ and  the
function $\chi$ used in it is no longer
identified with the BFKL kernel, but rather given by an `effective' $\chi$
function
\eqn\chiefdef{\eqalign{&\chi_{\rm eff}(\as,M)=\chi(\as,M) +\Delta
\chi(\as,M),\cr &\quad \Delta \chi(\as,M)=\as^2 \Delta\chi_1(M)+\dots,\cr }}
where $\chi(\as,M)$ is obtained letting
$\ash\to\as$ in the kernel of eq.~\xevolrun, and the correction term $\Delta
\chi$ to NLLx is given by
\eqn\deltachi{
\Delta \chi_{1}(M)=\beta_0
\frac{\chi_0''(M)\chi_0(M)} {2\chi_0'(M)}.    }

The problem with this perturbative approach is that the correction terms which
must be included in the effective
$\chi $ eq.~\chiefdef\ have an unphysical singularity: $\chi_0(M)$ has a
minimum at $M=\half$, so the denominator of
$\Delta \chi_{1}(M)$ vanishes, resulting in a simple pole in the NLLx
correction
$\Delta\chi_1$ eq.~\deltachi\ at $M=\half$. The NNLLx correction $\Delta\chi_2$
turns out to have a fourth--order pole, and
in fact at each extra order three extra powers of $(\chi'_0)^{-1}$ appear. As
explained in ref.~\sxap, this leads to a
perturbative instability: as a consequence of the singularity, the splitting
function $\Delta P_{ss}$ eq.~\psasy\
associated with the anomalous dimension $\Delta\gamma_{ss}$ behaves as
\eqn\splinst{{\Delta P_{ss}(\as,\xi)\over
P_s(\as,\xi)}\toinf{\xi}\left(\as\xi\right)^2.} 

This problem was completely
solved in ref.~\sxrun, where we have shown that the
$M=\half$ singularity and the  corresponding ones that appear at higher orders
in
$\as \beta_0$ are an artifact of the expansion, are not present in the
all--order solution and can thus be eliminated, thanks to the
possibility of matching the all-order solution to the perturbative one. The
corresponding resummation, which will be discussed in detail in the
next section, 
is obtained by explicitly solving eq.~\xevolrun\ for
$\chi(\ash,M)\sim \ash\chi_0$, with $\chi_0$ replaced by its quadratic
approximation near $M=\half$. As is well known, the
solution is given in terms of an Airy function. This solution is added to the
anomalous dimension with subtraction of the
NNLx expansion terms to avoid double counting. The
$\chi$ function dual to the resulting expression for $\gamma$ is now regular
near $M=\half$. 

In the sequel, we
will discuss in detail the form and the properties of the improved anomalous
dimension. In the next section we describe the
leading approximation, where as input we only use the one-loop perturbative
anomalous dimension $\as\gamma_0(N)$ and the
leading order BFKL kernel $\as\chi_0(M)$ but with the running coupling effects
taken into account. Subleading corrections, including those from
$\chi_1$, will then be discussed in section~4.

\newsec{The Improved Anomalous Dimension: First Iteration}

Assuming that one only knows $\gamma_0(N)$ and $\chi_0(M)$, we argue that the
optimal use that one can make of these inputs
is to write down for the improved anomalous dimension the following expression:
\eqn\leadimpr{\eqalign{
\gamma_I(\as,N) &= [\as\gamma_0(N)+\gamma_s(\smallfrac{\as}{N})
-\smallfrac{n_c\as}{\pi N}]+\cr &\qquad
+\left[\gamma_A(c_0,\as,N)-\half +\sqrt{\smallfrac{2}{\kappa_0\as}[N-\as
c_0]}+\quarter\beta_0\as\right]-\rm{mom.~sub.}}} The first line on
the right-hand side, within square brackets, is the usual
double-leading~\sxres\ expression for the improved anomalous
dimension at this level of accuracy, made up of the first order perturbative
term
$\as\gamma_0$ plus the power series of terms
$(\as/N)^n$ contained in $\gamma_s(\smallfrac{\as}{N})$, obtained
from $\chi_0$ using eq.~\lodual, with subtraction of the order
$\as$ term to avoid double counting ($c_A=n_c=3$).  In the second
line, the second pair of square brackets contain the running coupling
Airy term:
$\gamma_A(c_0,\as,N)$ is the Airy anomalous dimension, and the remaining terms
subtract the contributions to $\gamma_A(c_0,\as,N)$
which are already contained in
$\gamma_s$ and
$\gamma_0$ as we shall explain shortly.  Finally ``$\rm{mom.~sub.}$" is a
subleading subtraction  that ensures momentum
conservation
$\gamma_{I}(\as,N=1)=0$. We now specify $\gamma_A(c_0,\as,N)$ and  ``${\rm
mom.~sub.}$".

For $\gamma_A(c,\as,N)$~\foot{For convenience we only include $c$ (and not
$\kappa$, see below) among the explicit arguments of $\gamma_A$.} we
start from the Mellin transform of eq.~\xevolrun, namely~\sxrun\
\eqn\nevolrun{NG(N,M)=\chi(\ash,M) G(N,M)+F(M), } with kernel
\eqn\quadr{\chi_q(\ash,M)=\ash[c + {1\over 2} \chipp (M-M_s)^2]. } The
expression within squared brackets is the quadratic
approximation of
$\chi$ near its minimum $M=M_s$. At leading order,  $M_s$,
$c$ (the Lipatov exponent) and $\kappa$ are given
by:\eqnn\mszero\eqnn\czero\eqnn\cappazero$$\eqalignno{M_s&=M_s^{0}=
\half,&\mszero  \cr c&=c_0=\frac{4n_c}{\pi}\log{2}=2.64763...,&\czero\cr
\kappa&=\kappa_0=-\smallfrac{2 n_c}{\pi}\psi''(\half)=32.1406...,&\cappazero}$$
where
$\psi(x)$ is the digamma function. The Airy anomalous dimension is found from
the (Mellin transformed) solution
$G_q(N,t)$ of this equation:
\eqn\aithree{
\gamma_A(c,\as(t),N)\equiv\frac{d}{dt}\ln G_q(N,t)=M_s
 + \left(\frac{2\beta_0
N}{\chipp}\right)^{1/3}\frac{\Ai'[z(\as(t),N)]}{\Ai[z(\as(t),N)]}, } where
$\Ai(z)$ is the Airy function, which satisfies
\eqn\airy{
\Ai''(z)-z\Ai(z)=0, } with $\Ai(0)=3^{-2/3}/\Gamma(2/3)$,
\eqn\aitwo{ z(\as(t),N) \equiv
\left(\frac{2\beta_0 N}{\chipp}\right)^{1/3} \frac{1}{\beta_0}
\left[\frac{1}{\alpha_s(t)}-\frac{c}{N}\right]. }

Along the positive real axis, the Airy function is a positive definite,
monotonically decreasing function of its argument,
and it behaves asymptotically as
\eqn\airyasymp{
\Ai(z)= \half \pi^{-1/2}z^{-1/4}
\exp(-\smallfrac{2}{3}z^{3/2})\big(1+O(z^{-3/2})\big). } Thus the large--$z$
behaviour of $\gamma_A$, which corresponds to
$N$ large and/or $\as(t)$ small, is  given by
\eqn\aifive{\eqalign{ &\gamma_A(c,\as,N) =\gamma^A_s(\as/N)+ O\left({\as^2\over
N}\right)\cr &\qquad
\gamma^A_s(\as/N)=M_s - \sqrt{\frac{2}{\chipp}
\Big[\frac{N}{\as}-c\Big]}. }} The leading $\gamma^A_s$  behaviour coincides
with the `naive', i.e. fixed-coupling  dual
anomalous dimension: namely, the anomalous dimension which is found from
$\chi_q$ eq.~\quadr\ using the leading--order
duality relation~\dualdef.  Furthermore, the leading $O(\as^2/N)$
correction is  simply obtained from
eq.~\airyasymp:
 \eqn\rss{\as\gamma^A_{ss}(\as/N)=-\frac{1}{4}
\frac{\beta_0\as}{1-\frac{\as}{N}c},  } and it coincides  with the
next-to-leading running coupling correction
eq.~\deltag\ computed for the quadratic kernel eq.~\quadr. Subsequent terms in
the asymptotic expansion eq.~\aifive\
correspond to the further
$\as^2\gamma_{sss}$,~\dots\ corrections which are found when the perturbative
solution of the running coupling
eq.~\xevolrun\ is pursued to higher orders.  The asymptotic series is (Borel)
resummed~\sxrun\ by the Airy anomalous
dimension.  At leading order, the double-counting between the double-leading
terms and Airy anomalous dimension is removed
by subtracting the leading term of the asymptotic expansion eq.~\rss, as well
as the term
$\as\gamma^A_{ss}(0)=-\quarter\beta_0\as$, which is $O(\as)$ and thus must be
subtracted because it is already in $\as
\gamma_0$.

The momentum conservation subtraction ``${\rm mom.~sub.}$'' is an 
$O(\as^2)$ term
which can be defined as:
\eqn\momconn{{\rm mom.~sub.} = g(N) \bar {\gamma}_I(\as,1),} where $g(N)$ is a
weight function with $g(1)=1$, with no
singularities for
$N>0$ and $\bar{\gamma}_I(\as,N)$ is given by
\eqn\momcon{\gamma_I(\as,N)=\bar{\gamma}_I(\as,N)-{\rm mom.~sub.}.} We can take
$g(N)=1$ or, for instance,
\eqn\gN{g(N)=\frac{1+r}{N+r}} where $r$ is a positive real number. The possible
advantage of this choice is to have a
damped large $N$ behaviour. These different choices
are all equivalent up to subleading terms.

We now discuss the properties of the improved anomalous dimension in this
approximation. In the limit $\as
\rightarrow 0$ with arbitrary $N$, $\gamma_I(\as,N)$ reduces to
$\as\gamma_0(N)+O(\as^2)$: in particular as $N\to\infty$ 
the Airy term reduces to an
$O(\as^2)$ constant (the
$N\to\infty$ limit of the next term, $\gamma^A_{sss}$, in the expansion of
eqs.~\airyasymp--\rss\ up to an
$O(N^{-1})$ correction). For  $\as
\rightarrow 0$ with $\as/N$ fixed, $\gamma_I(\as,N)$  reduces to
$\gamma_{\rm DL-LO}=\as\gamma_0(N)+\gamma_s(\smallfrac{\as}{N})
-\smallfrac{n_c\as}{\pi N}$, i.e. the leading term of the
double-leading expansion. Thus the Airy term is subleading in both
limits. In spite of this, its role is very significant because of the
singularity structure of the different terms in
eq.~\leadimpr. In fact,
$\gamma_0(N)$ has a pole at $N=0$, $\gamma_s$ has a branch cut at
$N=\as c_0$, and
$\gamma_A$ has a pole at $N=N_0<\as c_0$, where $N_0$ is the position of the
rightmost zero of the Airy function. The
importance of the Airy term is that the square root term subtracted from
$\gamma_A$ cancels, within the stated accuracy, the branch cut of $\gamma_s$
at $N=\as c_0$ and replaces the
corresponding asymptotic behaviour at small $x$ with the much softer one from
$\gamma_A$.

\topinsert
\vbox{
\epsfxsize=10truecm
\centerline{\epsfbox{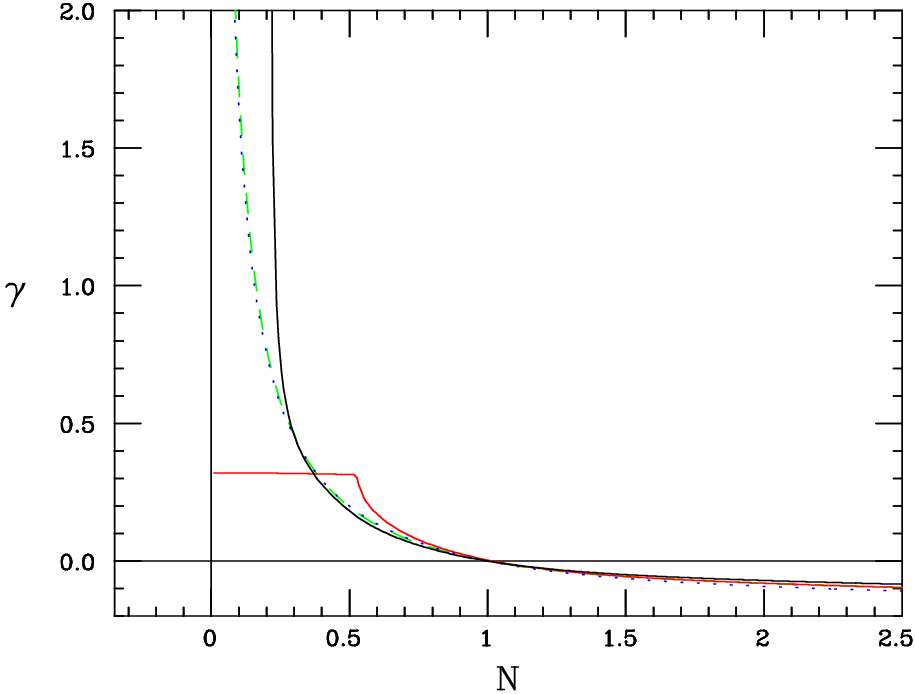}}
\hbox{
\vbox{\footnotefont\baselineskip6pt\narrower\noindent Figure 1: The improved
anomalous dimension $\gamma_I(\as,N)$ for
$c=c_0$, $\kappa=\kappa_0$, $M_s=1/2$ (i.e. the values in
eqs.~\czero,~\cappazero,~\mszero that are obtained from $\chi_0(M)$),
$g(N)=1$ and
$\as=0.2$ (solid), compared with
$\gamma_{\rm DL-LO}$ (the curve with a step) and GLAP LO and NLO (dashed and
dotted). }}\hskip1truecm}
\endinsert The effect of this replacement is clearly seen from
figure~1, 
where we
compare the plot of $\gamma_I$ (solid curve) as a function of
$N$ with those of $\gamma_{\rm DL-LO}$ (the curve with a step, starting at the cut
branch point) and of GLAP LO and NLO (which
are nearly superimposed).  The curves are computed with $c=c_0$,
$\kappa=\kappa_0$,  $M_s=1/2$
(i.e. the values in eqs.~\czero--~\mszero\ 
that are obtained from $\chi_0(M)$), to the momentum subtraction
with $g(N)=1$ eq.~\gN\ 
and $\as= 0.2$, a value which is roughly appropriate for the HERA
kinematic range. Since
$N_0$ is significantly smaller than $\as c_0$, the improved anomalous dimension
$\gamma_I$ remains close to the GLAP curves down to much smaller values of $N$
than $\gamma_{\rm DL-LO}$. It is only at $N\lsim
0.3$, where $\gamma_I>0.5$, that the improved anomalous dimension becomes
sizably different than the NLO perturbative
anomalous dimension (for the chosen value of $\as$ the pole of $\gamma_I$ is at
$N_0=0.211237$...while $\as c_0=0.529525$...).

\topinsert
\vbox{
\epsfxsize=10truecm
\centerline{\epsfbox{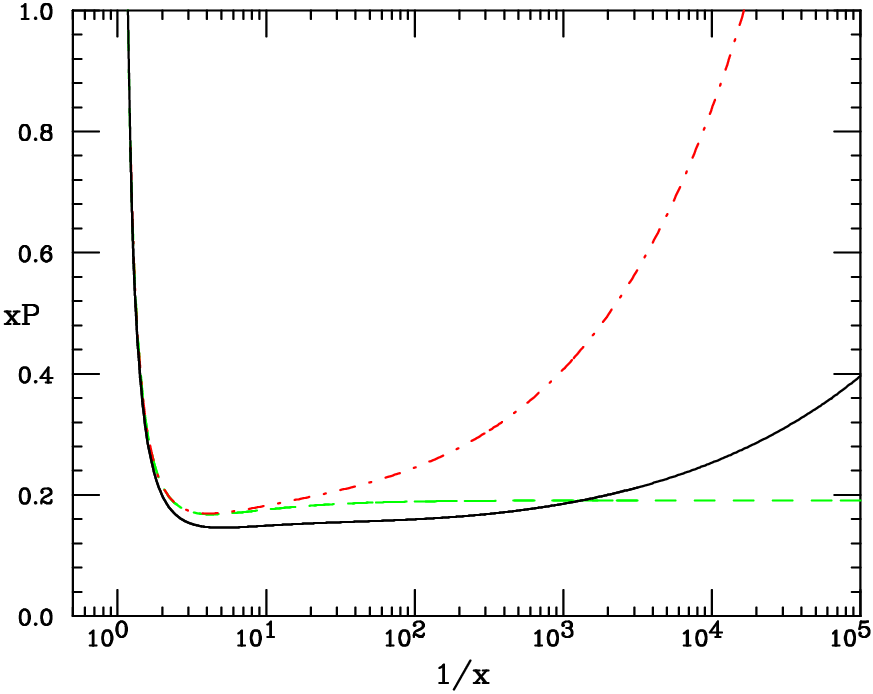}}
\hbox{
\vbox{\footnotefont\baselineskip6pt\narrower\noindent Figure 2: The improved
splitting function $P_{gg}(x)$ for
$c=c_0$, $\kappa=\kappa_0$, $M_s=1/2$ (i.e. the values in
eqs.~\czero,~\cappazero,~\mszero\ that are obtained from $\chi_0(M)$),
$g(N)=1$
and
$\as=0.2$ (solid), compared with that from $\gamma_{\rm DL-LO}$
(dot-dashed) 
and GLAP LO
(dashed). }}\hskip1truecm}
\endinsert The deviation of
$\gamma_I$ from GLAP at small values of $N$ has a moderate effect on the
splitting function. In fig.~2 we plot $P_{gg}(x)$
corresponding to $\gamma_I$ (dashed) compared to the perturbative LO (solid),
while the rising curve at small $x$
corresponds to $\gamma_{\rm DL-LO}$. This plot makes particularly clear that the
improvement obtained by implementing the Airy
asymptotics is really important at small $x$,  especially when we take into
account that the data seem to follow rather
closely the unresummed
perturbative evolution: the rapid small $x$ rise of the
splitting function is removed by the resummation, and in fact the
resummed splitting function closely follows the perturbative  one
down to $x\sim10^{-3}$.

The improvement of the  agreement with the leading order GLAP
anomalous dimension when the
double--leading anomalous dimension is supplemented
by running coupling resummation according to fig.~1 follows from three reasons.
First,  the location of the rightmost
singularity is lowered   from $\as c_0$ to $N_0$. Second,  the double--leading
branch cut is replaced by a pole, which is
the same type of singularity as in GLAP.  Finally, the residue of the pole in
$\gamma_A$ is more than an order of magnitude smaller
than the residue of the GLAP pole. Indeed, a straightforward calculation using
eq.~\aithree\ shows that
\eqnn\aires
$$\eqalignno{\gamma_A(\as(t),N)&={1\over N-N_0} {r_A}+ O[(N-N_0)^0]&\cr &r_A={3
\beta_0 N_0^2 \as(t)\over N_0+2 c_0
\as(t)}.&\aires}$$ When $\as=0.2$, $r_A=0.014...$ to be compared with
the value 
$\gamma_0= 3\as/\pi{1\over
N}+O[N^0]=0.191.../N+O[N^0]$
of the residue of the leading-order
GLAP pole.  The combination of these facts implies that  the
singularity in $\gamma_A$ only kicks in
rather late, \ie\ for rather small values of $N$. This is apparent in fig.~2,
where the rise $r_A x^{-N_0}$ due to the
Airy pole is only visible at very small $x$ on top of the constant behaviour
related to the GLAP pole. Due to momentum conservation, which fixes
the integral of $xP$, the rise must be compensated by  a small depletion of the
splitting function in the  intermediate  $x$ range, also visible in
the figure. While all these properties are generic, the quantitative
agreement of the resummed result with leading order GLAP is only good
when the intercept is
close to the leading order value $\as c_0$ [eq.~\czero] 
determined from the BFKL
kernel $\chi_0$. 
Running coupling resummation is the essential ingredient that
reconciles the BFKL intercept with the GLAP behaviour of the data
GLAP~\refs{\das,\dasnew}.

In our previous work in refs.~\refs{\sxres,\sxphen} we  had adopted a more
phenomenological approach in order to cope with
the fact that the rise of the singlet structure function implied by the Lipatov
exponent (corresponding to the anomalous
dimension with the cut at $\as c_0$ in fig.1) is too sharp in comparison to
what is seen in the data. Noting that the
correct asymptotic exponent is not reliably determined by the two known terms
of its perturbative expansion (in fact the
computed correction to the leading result is quite large), we had treated it as
a parameter to be fitted from the data. To
this effect, we interpreted the parameter
$\Delta
\lambda$ in eq.~\lambdadef, formally of order
$\as^2$, as a parameter to be fixed empirically. The lack of any indication of
a power-rise in the data suggests then that
$\Delta \lambda$ should be negative and  possibly numerically as large as, say,
$O(\as)$. In leading order, the corresponding improved
anomalous dimension becomes
\eqn\leadold{\eqalign{
\gamma_{Iold}(\as,N) &= [\as\gamma_0(N)+\gamma_s(\smallfrac{\as}{N-\Delta
\lambda}) -\smallfrac{n_c\as}{\pi N}]-\rm{mom.~sub.}}} The shift $N\rightarrow
(N-\Delta \lambda)$ in $\gamma_s$ provides
the change from $x^{-\lambda_0}$ to $x^{-\lambda}~=~x^{-\lambda_0} e^{\Delta
\lambda
\xi}$. As discussed in ref.~\sxres, there is an ambiguity in the subtraction
$n_c\as/(\pi N)$ because there too we could make the replacement $N\rightarrow
(N-\Delta \lambda)$.   In eq.~\leadold\ we
have adopted the subtraction with $N$, denoted as R-subtraction in
ref.~\sxphen,  because it is more suitable for the
present purpose of comparing the old approach with $\gamma_I$ in
eq.~\leadimpr, as we shall see shortly.

What we have accomplished in this section is to show that the resummation
through the Airy asymptotics  of higher--order
contributions to the anomalous dimension related to the  running coupling
corrections is by itself sufficient to produce
the required softening of the behaviour at small $x$. Thus it is no
longer  necessary to introduce
$\Delta
\lambda$ by hand as a free parameter (at least at leading accuracy). In fact,
we had already found in ref.~\sxres\ that
the value of
$\Delta\lambda$ that provides the best matching to GLAP is exactly such that
$\lambda=\as c_0+\Delta \lambda\sim 0.21$ (for
$\as\sim0.2$) in remarkable agreement with the value $N_0=0.211237$ determined
here from the Airy asymptotics. The fact
that the data are in good agreement with unresummed perturbative evolution
suggests that the true all-order value of
$\lambda$ is likely to be quite close to the leading--order value, so $\Delta
\lambda$ is presumably rather small and
there is perhaps no need to introduce it as a free parameter in order to fit
the data. 

The
sensitivity of our new approach to the value at the minimum $\lambda=\as c= \as
c_0+\Delta \lambda$ can be demonstrated by going back to
eq.~\leadimpr, replacing $N$ by $N-\Delta\lambda$ in $\gamma_s$ and $c_0$ by
$c$ in the Airy term (including the square
root subtraction):
\eqn\leadimprdel{\eqalign{
\gamma_I(\as,N,c) &= [\as\gamma_0(N)+\gamma_s(\smallfrac{\as}{N-\Delta \lambda})
-\smallfrac{n_c\as}{\pi N}]+\cr &\qquad
+\gamma_A(c,\as,N)-\half +\sqrt{\smallfrac{2}{\kappa_0\as}[N-\as
c]}+\quarter\beta_0\as-\rm{mom.~sub.}}}
Note that we have not performed the $\Delta \lambda$ shift in
the double-counting subtraction term,
corresponding to the R-subtraction of ref.~\sxphen. This is necessary,
otherwise the cancellation of singularities at $N=\as c$ between
the double-leading anomalous dimension and the Airy term would be spoiled.
 The resulting improved anomalous dimension for $\as=0.2$ and $\as c=
0.4,~0.529525...$ (the BFKL value,
see eq.~\czero), and
$0.6$ are shown in figure 3. The corresponding splitting functions are shown in
figure 4. We see that changing $c$ in this
range does not alter the overall trend of the anomalous dimension. However the
impact on the splitting function at small $x$ is
quite noticeable and, in fact, the observed approximate validity of the
perturbative fits suggests that the viable range of
$\Delta \lambda$ is  inside the interval presented in figure 4. This
possibility of varying
$c$ in the leading term could be adopted as a phenomenological way to optimize
the leading formula on the data. We have also
checked that moderate variations of
$\kappa_0$ and $M_s$ eqs.~\mszero,\cappazero, do not alter the general picture.
\topinsert
\vbox{
\epsfxsize=10truecm
\centerline{\epsfbox{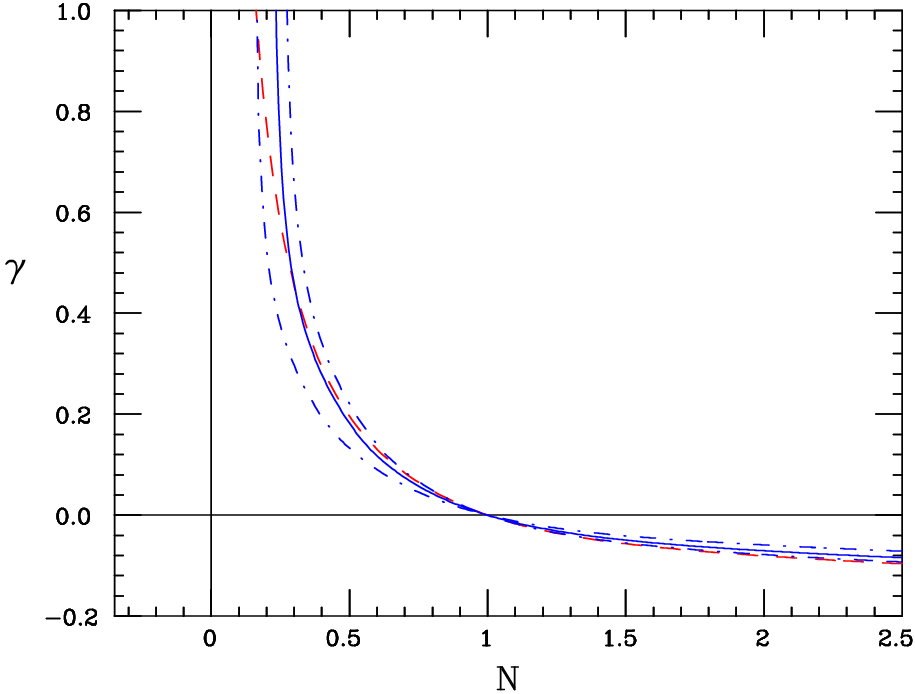}}
\hbox{
\vbox{\footnotefont\baselineskip6pt\narrower\noindent Figure 3: The effect on
the improved anomalous dimension of varying the
value of $\chi$ at the minimum: $\lambda=\as c= \as c_0+\Delta \lambda$, as
described in the text, with $\as=0.2$ and $\as c=
0.4, 0.529525..., 0.6$. The dashed curve is the LO GLAP curve, the solid
curve corresponds to the BFKL value $\as c=
0.529525...$ while the upper and lower (at small $N$) dot-dashed 
curves refer to
$\as c=
0.6,~0.4$, respectively. }}\hskip1truecm}
\endinsert
\topinsert
\vbox{
\epsfxsize=10truecm
\centerline{\epsfbox{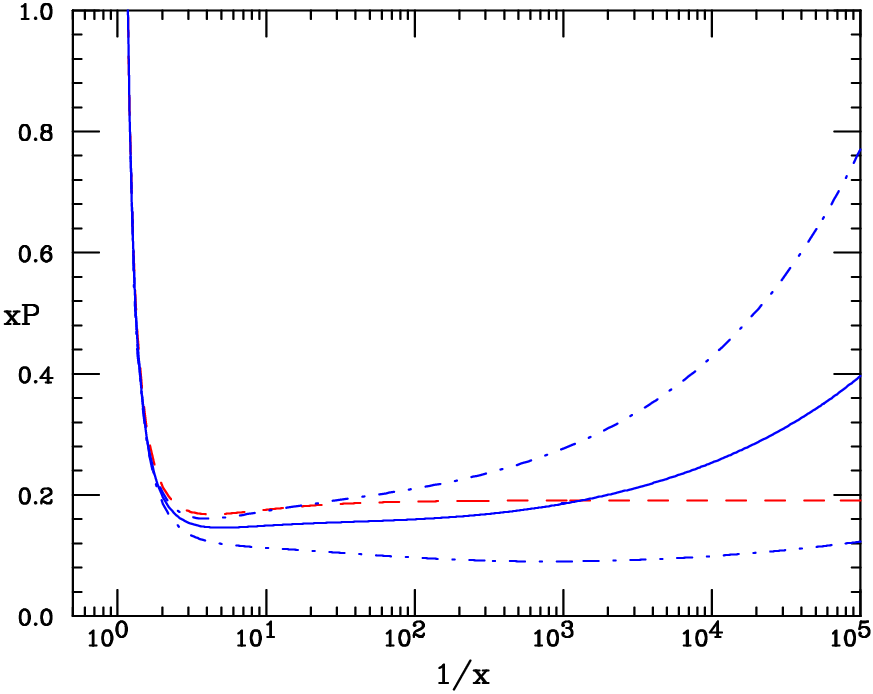}}
\hbox{
\vbox{\footnotefont\baselineskip6pt\narrower\noindent Figure 4: The effect on
the improved splitting function of varying the
value of $\chi$ at the minimum: $\lambda=\as c= \as c_0+\Delta \lambda$, as
described in the text, with $\as=0.2$ and $\as c=
0.4, 0.529525..., 0.6$. The dashed curve is the LO GLAP curve, the solid
curve corresponds to the BFKL value $\as c=
0.529525...$ (same as the solid curve of Fig.~1)
while the upper and lower (at large $1/x$) dotdashed 
curves refer to
$\as c=
0.6,~0.4$, respectively. }}\hskip1truecm}
\endinsert

Since the reference perturbative fits are based on the NLO GLAP it may be
desirable to modify our improved anomalous
dimension in such a way that at small $\as$ with $\as/N$ fixed it reduces to
NLO GLAP given by eq.~\gammadef:
$\gamma(N,\alpha_s)=\alpha_s
\gamma_0(N)~+~\alpha_s^2 \gamma_1(N)$. This can readily be obtained by
modifying eq. ~\leadimpr\ in the following
way:
\eqn\leadimprnl{\eqalign{
\gamma_I^{NL}(\as, N) &= [\as\gamma_0(N)+ \as^2 \gamma_1(N) +
\gamma_s(\smallfrac{\as}{N}) -\smallfrac{n_c\as}{\pi N}]+\cr &\qquad
+\gamma_A(c_0,\as,N)-\half +\sqrt{\smallfrac{2}{\kappa_0\as}[N-\as
c_0]}\cr&\qquad\qquad
+\quarter\beta_0\as(1+\frac{\as}{N} c_0)-\rm{mom.~sub.}}}
Here we used the fact that in the expansion  of
$\gamma_s(\as/N)$ in powers of its argument the quadratic term is absent, so
that it is sufficient to only add a
subtraction to the Airy term  in order to guarantee that in
the perturbative limit the NLO GLAP anomalous
dimension is exactly recovered. This subtraction (the first term in
the third line of eq.~\leadimprnl) is obtained by
expanding $\as\gamma_{ss}^A$ eq.~\rss\ to order $\as^2$ at fixed $N$. This formula improves the NLO GLAP with the
leading terms of the BFKL corrections with running
coupling. The result is shown in figure 5 and compared to the  GLAP
anomalous dimension and the resummed LO result of
Figure~1. It appears that, thanks to the extra next-to-leading
subtraction in eq.~\leadimprnl, 
the effect of the resummation is further reduced, in particular
correcting the slight distortion of the LO resummed result in the
$0.5\lsim N\lsim 1.5$ region seen in the figure
(dotdashed curve).
\topinsert
\vbox{
\epsfxsize=10truecm
\centerline{\epsfbox{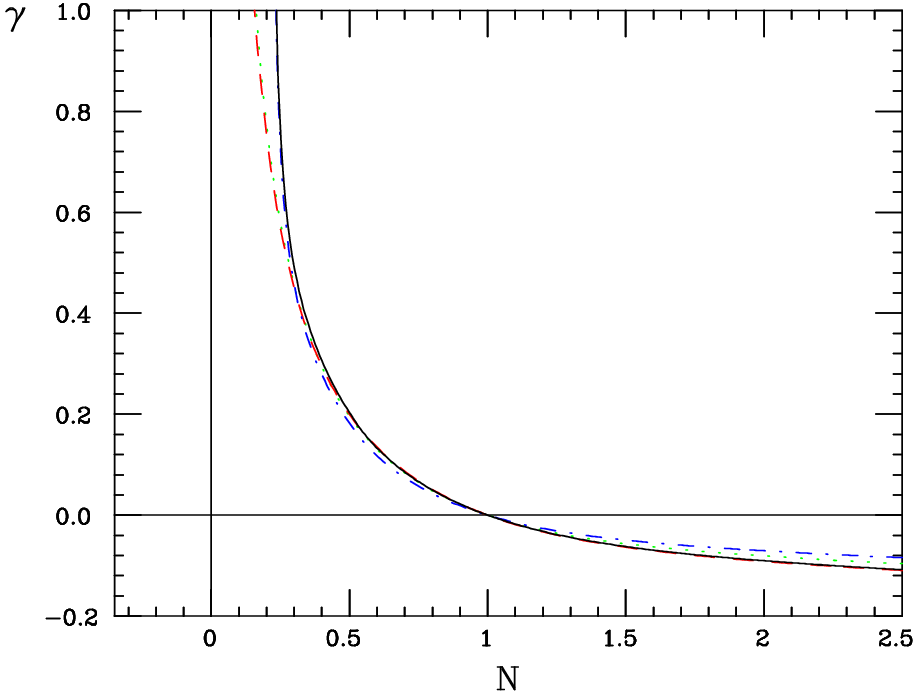}}
\hbox{
\vbox{\footnotefont\baselineskip6pt\narrower\noindent Figure 5: The improved
anomalous dimension $\gamma_I^{NL}$
eq.~\leadimprnl (solid), LO (dotted)
and NLO GLAP (dashed), which is the
corresponding perturbative limit in this case, for
$c=c_0$, $\kappa=\kappa_0$, $M_s=1/2$ (i.e. the values in
eqs.~\czero,~\cappazero,~\mszero\ that are obtained from $\chi_0(M)$),
$g(N)=1$
and
$\as=0.2$. The resummed LO curve (solid curve of Figure 1) 
is also shown for comparison (dotdashed).
}}\hskip1truecm}
\endinsert

In conclusion we have shown that, given the tendency of the data to closely
follow the GLAP evolution down to the smallest
values of $N$ accessible to HERA, the most effective way to implement the
information contained in the leading BFKL kernel
$\chi_0$ is to use duality, the double-leading prescription (that improves
$\chi_0$ near $M=0$ in agreement with momentum conservation), and the Airy
asymptotics to get just the right amount of
softening of the small $x$ exponent. In the next section we will show that the
non leading corrections, including those
arising from $\chi_1$, do not spoil this overall picture.

\newsec{Perturbative expansion of the resummed anomalous dimension}

In the previous section we have seen that at leading order the resummed
anomalous dimension eq.~\leadimpr\ and the GLAP
anomalous dimension are in very good agreement. 
However, this result is phenomenologically significant only if the
resummed anomalous dimension eq.~\leadimpr\
is stable upon higher order corrections. To investigate whether this is the
case we must discuss how eq.~\leadimpr\ can be
viewed as the leading order of a resummed perturbative expansion of the
anomalous dimension.  As suggested in ref.~\sxrun,
this can be done  by using the resummation of running coupling effects by means
of the Airy asymptotics to improve the
expansion eq.~\sxexp\ of the anomalous dimension in powers of $\as$ at fixed
$\as/N$, and then  combining this  improved
expansion with the standard GLAP anomalous dimension, to obtain a
running-coupling improvement of the double leading
expansion, thereby generalizing to all orders eq.~\leadimpr.

Hence, we must 
consider the running-coupling resummation of the small-$x$ expansion
of the anomalous dimension
eq.~\sxexp\ beyond leading order.  Higher order
corrections $\chi_i$ to the kernel $\chi_0$ will change in general the values
of  all the parameters~\mszero--\cappazero\
which characterize the quadratic approximation to $\chi$ near its minimum.
Whenever the value of $\chi$ at its minimum
differs from the leading--order value, the perturbative expansion of the kernel
$\chi$ must be reorganized if we wish to obtain a stable expansion of the
splitting function, \ie\ such that the ratios
$P_{ss}/P_{s}$,
$P_{sss}/P_{s}$,~\dots\ all remain finite as $\xi\to\infty$. A necessary and
sufficient condition for this to be the case
is~\refs{\sxap,\sxres}  that to each order a constant is subtracted  from
$\chi_i$ and added back into $\chi_0$, in such a
way that the subtracted  $\chi_i$ no longer  lead to a shift in the minimum  of
$\chi_0$. At leading order, this amounts to performing the shift
$N\rightarrow N-\Delta \lambda$ as in eq.~\leadimprdel.

In view of a running coupling resummation, however, it is convenient to include
in $\chi_0$ the full leading asymptotic
behaviour, by subtracting not only a constant, but also the leading linear and
quadratic terms in the Taylor expansion of
$\chi$ about its all--order minimum.  Namely, assume that the all-order $\chi$
has a  minimum in the vicinity of which it
can be approximated by the quadratic expression of $\chi_q$, as  given in
eq.~\quadr, where now
\eqnn\msx\eqnn\cx\eqnn\cappax$$\eqalignno{M_s&=M_s^{0}+\as M_s^{(1)}+\dots
,&\msx  \cr
 c&=c_0+\as c_1+\dots,&\cx\cr
\kappa&=\kappa_0+\as\kappa_1+\dots.&\cappax}$$ We then reorganize  the
expansion of $\chi$ eq.~\chidef\ as follows:
\eqn\chiresum{\eqalign{\chi(M,\as)&= \as \chi^R_0 +\as^2
\chi^R_1+\dots;   \cr 
\chi^R_0&=\chi_0+\chi_q-\chi_q^{(0)};\cr 
\chi^R_i&=\chi_i-\chi_q^{(i)},\qquad i>0;\cr 
\chi_q^{(i)}&\equiv \chi_i(M_s)+(M-M_s)\, \chi'_i(M_s) +\half
(M-M_s)^2\chi''_i(M_s), \qquad i\ge 0.\cr }} To any finite
order, the  expansion eq.~\chiresum\ of $\chi$  differs from the standard one
eq.~\chidef\ by subleading terms:
\eqn\resdif{\sum_{i=0}^n \as^{i+1} \chi_i(M)=
\sum_{i=0}^n \as^{i+1} \chi_R^{(i)}(M)+O[\as^{n+1}].} Therefore, eq.~\chiresum\
is a {\it bona fide} resummed version of the
original expansion. We may then use the resummed expansion to determine order
by order $\gamma_s^R$, $\gamma_{ss}^R$ and
so forth, through perturbative duality with the suitable running coupling
corrections. In particular, the next-to-leading
order running coupling contribution to $\gamma_{ss}^R$ is still given by
eq.~\deltag, but  with $\chi_0$ replaced by its
resummed counterpart $\chi_0^R$ eq.~\chiresum. This leads to  a resummed
expansion of $\gamma$ in powers of $\as$ at fixed
$\as/N$ which, to any finite order $n$, differs by the standard expansion
eq.~\sxexp\ by subleading $O(\as^{n+1})$ terms.
Because, by construction, $\chi_R^{(i)}$ do not shift the minimum of
$\chi_R^{(0)}$, the ensuing resummed expansion of $\gamma$ is stable, just like
the expansion discussed in our previous
work refs.~\refs{\sxap,\sxres,\sxphen}. 

This new reorganized expansion
differs 
from  that discussed in refs.~\refs{\sxap,\sxres} because now not
only the value of $\chi$ at the minimum, but also its curvature
is reshuffled into the leading order, and furthermore, the expansion is
performed about the all--order minimum $M_s$ instead of the
leading order minimum $M=\half$. The leading order resummed 
$\chi_0^R$ has in general not only a value at the minimum 
$c\not=c_0$, like the leading order resummed $\tilde \chi_0$ of
refs.~\refs{\sxap,\sxres}  but also
a minimum at $M_s\not=\half$ and
curvature $\kappa\not=\kappa_0$. Only if
$\kappa=\kappa_0$ and $M_s=\half$, then $\chi_0^R$ would be of the same
form as the leading order resummed $\tilde \chi_0$ of
refs.~\refs{\sxap,\sxres}. Note, however, that even so, the
subsequent terms $\chi_i^R$ with $i>1$ in the expansion of $\chi$ would
not be the same as the corresponding $\tilde \chi_i$ of
refs.~\refs{\sxap,\sxres} (compare eq.~\chiresum).

The resummed $\gamma$ at any desired order can now be combined with the Airy
anomalous dimension $\gamma_A$ determined
from the quadratic kernel with generalized~\msx-\cappax\ values of the
parameters. However, if the location of the minimum
$M_s$ differs from its leading order value,
$M_s\not=\half$, already at next-to-leading
order this is no longer sufficient to resum
all singularities related to the running coupling term $\Delta \gamma_{ss}$
eq.~\deltag. Indeed, if $M_s\not=\half$,
$\chi_0^R$ is no longer symmetric about $M=M_s$ and therefore its Taylor
expansion about
$M=M_s$ also contains in general a cubic term.  It is straightforward to see
that in such case $\Delta \gamma_{ss}$
acquires a new singularity.

Consider the case of a generic kernel $\chi$, with a minimum at $M=M_s$, and
containing terms up to fourth order:
\eqn\chiquart{\chi^{(4)}(\hat\as,M)=\chi_q+\hat\alpha \left[{\delta\over
3!}(M-M_s)^3+{\epsilon \over 4!}(M-M_s)^4\right].}
Expanding the anomalous dimension $\gamma_s^{(4)}(\as/N)$  which is dual
(eq.~\dualdef) to this
$\chi^{(4)}$ about its rightmost singularity $\as/N=c$ yields
\eqn\gamquart{\eqalign{&\gamma_s^{(4)}(\as/N)= M_s-\sqrt{{2\over\kappa}\left(
{N\over\as}-c\right)}\Bigg\{1-{\delta\over\kappa3!}\sqrt{{2\over\kappa}\left(
{N\over\as}-c\right)}\cr&\qquad\qquad\qquad\qquad
-{\epsilon\over\kappa4!}\left[{2\over\kappa}\left(
{N\over\as}-c\right)\right] +O\left[\delta^2\right]+O\left[\epsilon^2\right]
\Bigg\},\cr}} so higher order terms in the expansion of $\chi$ in powers of
$M-M_s$ correspond to higher order
contributions in the expansion of its dual anomalous dimension in powers of
$\sqrt{N/\as-c}$.
 Substituting $\gamma_s^{(4)}$ (eq.~\gamquart) in the expression eq.~\deltag\
of the running coupling contribution to
$\gamma_{ss}$ leads to
\eqn\cubsing{\gamma_{ss}^{(4)}\equiv\frac{\chi_0''(\gamma^{(4)}_{s})
\chi_0(\gamma_{s}^{(4)})}  {2\chi_0'^2(\gamma_{s}^{(4)})}=
\gamma^A_{ss}+\frac{\beta_0\delta}{24}\left({2\over\kappa}\right)^{3/2}
\sqrt{N\over\as}{1\over\sqrt{1-{\as\over N} c}}+ O\left[\left({N\over\as}-c
\right)^0\right],} where $\gamma^A_{ss}$ is the running coupling contribution
eq.~\rss\ which is resummed by the Airy
anomalous dimension. The cubic term induces the square-root singularity
proportional to $\delta$ in eq.~\cubsing, which
must also be resummed. Because the singularity is linear in $\delta$, this
resummation can be performed by treating the
cubic term as a leading-order perturbation. Note that the quartic term does not
induce any further singularities to this
order.

In order to resum running-coupling singularities which are associated to the
cubic term in the expansion of $\chi$ about
its minimum, we  determine the (Mellin transformed)  solution $G^{(3)}(N,t)$ to
eq.~\xevolrun\ when the cubic term is also
included in the kernel:
\eqn\cubsol{G^{(3)}(N,t)=\left({2 N \beta_0\over\kappa}\right)^{1/3}
\exp{M_s\over\beta_0\as(t)}
\exp-\left\{{\delta\over12\kappa}\left({2N\beta_0\over\kappa}\right)^{1/3}
{d^4\over dz^4}\right\}
Ai\left[z(\as(t),N)\right]. } Using the identity $Ai''(z)=z Ai(z)$ and
expanding in powers of
$\delta$ we get the anomalous dimension
\eqn\cubairy{\eqalign{&\gamma_A^{(3)}(\as(t),N)\equiv\frac{d}{dt}\ln
G_q^{(3)}(N,t)\cr &\quad\qquad=\gamma_A
-{\delta\over12}\left({2\over\kappa}\right)^{5/3}
\left({\beta_0 N}\right)^{2/3}\left[2 z(\as(t),N) -\left({Ai'(z(\as(t),N))\over
Ai(z(\as(t),N))}\right)^2\right]+O[\delta^2],}} where $\gamma_A$ is given by
eq.~\aithree.

The last term inside the square bracket in eq.~\cubairy\ has a double pole at
$N=N_0$, where $Ai(z(\as,N))$ vanishes. This
is due to the fact that the location of the zero of $G^{(3)}(N,t)$ eq.~\cubsol\
is shifted away from $N_0$  by the
$\delta$ correction, and therefore the pole in the anomalous dimension is also
shifted. To first order in $\delta$, the
zero moves from $N_0$ to
\eqn\shiftzer{N_0^\delta= N_0+{\delta\over6\kappa}r_A,} with $r_A$ given by
eq.~\aires. Because $r_A$ is $O(\as)$, the
ensuing  correction to the asymptotic behaviour is
subleading~\sxrun.\foot{Note that the explicit expressions of the
correction to the solution and the location of the pole in the  anomalous
dimension given in eq.~(5.2) and
(5.3) of ref.~\sxrun\ are incorrect, and should be respectively 
replaced by eq.~\cubsol\ and
eq.~\shiftzer. The main conclusion that the
modification of the asymptotic behaviour due to cubic and higher corrections is
subleading remains however unaffected.} It
follows that the linearized expression eq.~\cubairy\ of the $\delta$ correction
to the anomalous dimension has a double
pole in $N_0$, and is correct provided if $N-N_0$ is not too small, i.e. if
$x$ is not too small. For the typical values
of the parameters ($\as c\sim 0.4$),
$N_0^\delta-N_0\sim 10^{-2}$, so the correction to the asymptotic behaviour
$x^{-N_0}\rightarrow x^{-N_0^\delta}$ is approximately
$\sim10\%$ when
$x\sim10^{-5}$, and only becomes of order one when $x$ is extremely
small. 
Hence,
even though the linearized anomalous dimension
eq.~\cubairy\ does not reproduce the correct leading singularity of the full
anomalous dimension determined from the
solution eq.~\cubsol, it is a good approximation to it for all practical
purposes.

We can now determine the expansion of $\gamma_A^{(3)}$ in powers of $\as$ at
fixed $\as/N$
\eqn\threeasymp{\gamma_A^{(3)}={\gamma_{A,\,s}^{(3)}}_s(\as/N)+
\as{\gamma_{A,\,ss}^{(3)}}(\as/N)+O(\as^2),} by using the asymptotic expansion
eq.~\airyasymp:
\eqnn\threes\eqnn\threess$$\eqalignno{
\gamma_{A,\,s}^{(3)}(\as/N)=&\gamma^A_s-{\delta\over12}
\left({2\over\kappa}\right)^2\left({N\over\as}-c\right), &\threes\cr
\gamma_{A,\,ss}^{(3)}(\as/N)=&\gamma^A_{ss}+ {\beta_0\delta\over24}\left(
{2\over\kappa}\right)^{3/2}{N/\as\over
\sqrt{{N\over\as}-c}} ,&\threess }$$ where $\gamma^A_{s}$ and $\gamma^A_{ss}$
are given by eq.~\aifive\ and eq.~\rss\
respectively. This verifies explicitly that the $O(\delta)$ running coupling
contributions are correctly resummed by the
correction term in eq.~\cubairy: the leading-order term~\threes\ in the
expansion eq.~\threeasymp\ matches the naive
(fixed coupling) dual eq.~\gamquart, while the first subleading correction
matches the next-to-leading running coupling
contribution eq.~\cubsing.

We can thus write down the next-to-leading version of  the resummed expression
eq.~\leadimpr:
\eqn\gamresexp{
\gamma(\as,N)=\gamma_I(\as,N)+\as \gamma_{II}(\as,N)+O(\as^2)-\rm{mom.~sub.}.}
The leading-order term $\gamma_I$ is now
given by
\eqn\leadimprres{\eqalign{
\gamma_I(\as,N) &= \left[\as\gamma_0(N)+\gamma_s^R(\smallfrac{\as}{N})
-\smallfrac{n_c\as}{\pi N}\right]+\cr &\qquad
+\gamma_A^{(3)}(\as,N)-M_s +\sqrt{\smallfrac{2}{\chipp\as}[N-\as
c]}-\quarter\beta_0\as+{\delta\over12}
\left({2\over\kappa}\right)^2\left({N\over\as}-c\right),}} where  $\gamma_s^R$
is determined from
$\chi_0^R$ eq.~\chiresum\ using duality eq.~\dualdef,
$\gamma_A^{(3)}$ is  given by eq.~\cubairy, and the same values of the
parameters $M_s$, $\kappa$ and $c$ are used in
these two contributions. Because, as discussed above, if
$\kappa=\kappa_0$ and $M_s=\half$ the resummed $\chi_0^R$
eq.~\chiresum\ reduces to the resummed $\tilde \chi_0$ of
refs.~\refs{\sxap,\sxres}, it follows that in such case $\gamma_s^R(\as/N)$
coincides with $\gamma_s[\as/(N-\Delta \lambda)]$ with $\Delta
\lambda=(c-c_0)/\as$. 
 In
other words, this is the same as eq.~\leadimpr, but with generic values of the
parameters which characterize the behaviour of
$\chi$ about its minimum, eq.~\quadr, and the  cubic contributions  to the Airy
anomalous dimension accordingly included. 

The
next-to-leading correction is
\eqn\nlimprres{\eqalign{
\gamma_{II}(\as,N) = &\left[\as\gamma_1(N)+\gamma_{ss}^R\left({\as\over
N}\right) -e_0-\as\left({e_1\over N}+{e_2\over
N^2}\right)\right]\cr &\qquad\qquad +\as\quarter{\beta_0 c\over N -\as c}
-{\beta_0\delta\over24}\left(
{2\over\kappa}\right)^{3/2}{N/\as\over \sqrt{{N\over\as}-c}}.}} Here, the term
in square brackets is the analogue of the 
usual~\refs{\sxres,\sxphen}  next-to-leading order term in the double-leading
expansion (but determined from the new resummed $\chi_1^R$
eq.~\chiresum) while the last two terms are the
$O(\as^2/N)$ terms eq.~\cubsing\ in the expansion of the Airy anomalous
dimension.

Let us now discuss the properties of this next-to-leading order expression.
First, it is easy to see that the anomalous
dimension eq.~\gamresexp\ only differs from the standard next-to-leading
double-leading expansion of
refs.~\refs{\sxres-\sxphen}  by sub-subleading terms. Indeed, we have seen that
because of eq.~\resdif\ the reorganized
expansion of the anomalous dimension in terms of $\gamma_s^R$,
$\gamma_{ss}^R$~\dots\ differs from the usual one by
subleading terms. Furthermore, terms of order up to $\as^2$ in the expansion
eq.~\threeasymp\ of the Airy anomalous
dimension are explicitly subtracted: note in particular that the sum of the
order $\beta_0 $ term in $\gamma_I$ and the
order $\beta_0 c$ term in $\gamma_{II}$ exactly matches the subleading term
eq.~\rss\ in the expansion of
$\gamma_A$.

The small $N$ behaviour of both $\gamma_{I}$ and
$\gamma_{II}$  is controlled by the pole of $\gamma_A^{(3)} $at $N=N_0$, whose
residue receives a
$O(\delta)$ correction (compare eq.~\cubairy), while singularities at
$N=\as c$ in the  expansion eq.~\sxexp\ of the anomalous dimension are
subtracted to the stated accuracy.  As it is
apparent from eq.~\gamquart, terms of increasingly higher order in the
expansion of
$\chi$ about its minimum correspond to singular contributions to increasingly
higher order derivatives of $\gamma$ at the
point $N=\as c$. Namely, the quadratic term in $\chi$  leads to a singularity
in the first derivative of $\gamma$, the
cubic and quartic term to a singularity in the second derivative, and so on.
Hence, at the leading ($\gamma_s$) level, the
square-root cut at $N=\as c$ is subtracted  so that the anomalous dimension is
continuous and has a continuous first
derivative at $N=\as c$, though it has discontinuous second derivative, which
would  be made continuous by  including the
quartic $\epsilon$ term eq.~\gamquart\ in the Airy resummation. 

At the
next-to-leading  ($\gamma_{ss}$) level, thanks to
the reorganization of the expansion eq.~\chiresum\ there are no singularities
in the fixed-coupling dual anomalous
dimension nor in its first derivative.  There are however singularities related
to the running coupling term eq.~\deltag:
the pole (related to the quadratic contribution) and square-root singularity
(related to the cubic contribution) are
subtracted, while the first derivative has a discontinuity that would be
removed by including cubic corrections
eq.~\cubairy\ to $O(\delta^2)$.

At large $N$, the $\sqrt N$ growth of $\gamma_A$ is canceled by the
$\gamma^A_s$ subtraction, while the remaining constant
large $N$ limit  is canceled by the $\gamma^A_{ss}$ subtraction,  leaving a
$1/\sqrt{N}$ drop.  The linear  and $\sqrt N$
growth of $\gamma_A^{(3)}$  are canceled by the $\gamma_{A,\,s}^{(3)}$ and
$\gamma_{A,\,ss}^{(3)}$ subtractions
respectively, leaving an $O(\as^2)$ constant:
\eqn\threasconst{\lim_{N\to\infty}
\left(\gamma_{A}^{(3)}-\gamma_{A,\,s}^{(3)}-\gamma_{A,\,ss}^{(3)}\right)=
(\as\beta_0)^2 {\delta\over 24 \kappa}.} This
constant may be included in the momentum subtraction term. This leaves then an
$O(1/\sqrt N)$ drop. Therefore, at large
$N$ the  next-to-leading resummed anomalous dimension eq.~\gamresexp\ coincides
with the standard GLAP anomalous
dimension, since the subtracted  small $x$ eq.~\sxexp\ terms drop at least as
$1/N$ and the subtracted Airy terms drop at
least as $1/\sqrt N$, while the GLAP anomalous dimension, as well known, grows
(in modulus) as a power of
$\ln N$. This property is preserved if the momentum subtraction is performed by
subtracting the constant eq.~\threasconst,
and then subtracting the remaining momentum non-conserving terms through
eqs.~\momconn-\gN. If momentum is instead
enforced by simply subtracting a constant, the asymptotic  large $N$ resummed
anomalous dimension will differ from the
GLAP one by a fixed $O(\as^2)$ constant.

We conclude that to next-to-leading order the analytic properties of
the leading order result
eq.~\leadimpr\ are preserved: the branch cut at
$N=\as c$ is subtracted from the small $x$ anomalous dimension
$\gamma_s+\as\gamma_{ss}$ and replaced by
the Airy pole, while the large $N$ behaviour of the perturbative
anomalous dimension $\gamma_0+\as\gamma_1$ is preserved.

We can now check that the next-to-leading order anomalous dimension
eq.~\gamresexp\ differs from the leading order one
eq.~\leadimpr\ displayed in figure~1 by a small correction. We expect this on the
grounds that at large $N$ they reduce to
the corresponding GLAP anomalous dimension, while at small $N$ for $N\gsim
N_0^\delta$  they are controlled by the Airy
pole, whose location depends very little on the input value of $c$. At
next-to-leading order, however, we cannot use the
perturbative values of the parameters $c$, $\kappa$ and $M_s$, because the
next-to-leading order $\chi$ does not have a
minimum. This is due to the fact that, whereas the small $M\sim 0$ drop of
$\chi_1$ is removed by the double-leading
resummation, fixed by momentum conservation, the large $M\sim1$ drop of
$\chi_1$ can only be removed through model-dependent assumptions, such as those
of refs.~\refs{\salam,\ciaf}. Because the
series of small $M$ poles has alternating signs, this problem would disappear
at the next-to-next-to-leading-log level,
while at this order $c$, $\kappa$ and $M_s$ must be treated as free parameters,
upon the assumption that the minimum is
restored by higher order corrections.

\topinsert
\vbox{
\epsfxsize=10truecm
\centerline{\epsfbox{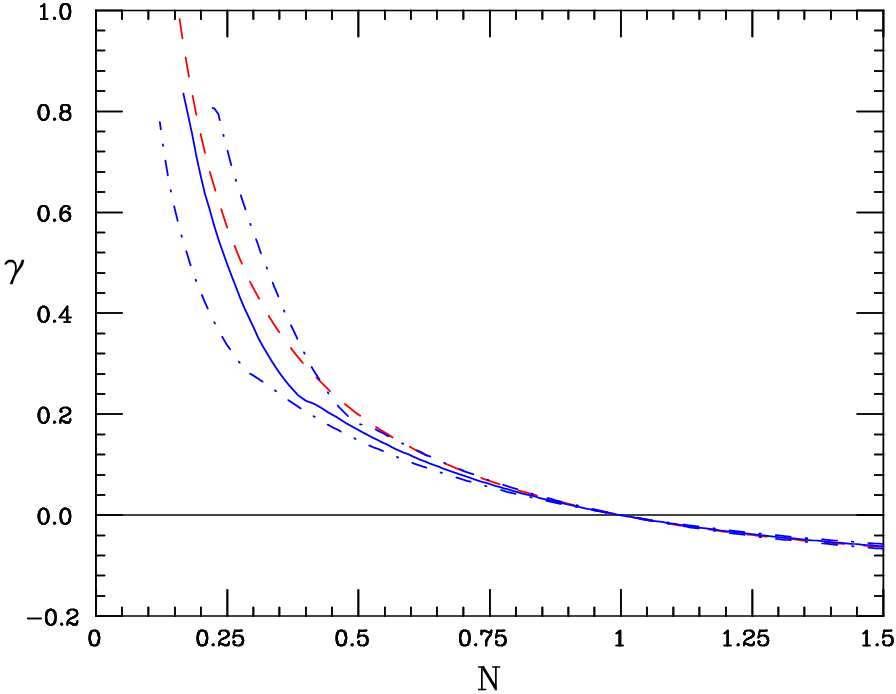}}
\hbox{
\vbox{\footnotefont\baselineskip6pt\narrower\noindent Figure 6: The improved
anomalous dimension eq.~\gamresexp\ for
$\as c=0.3$ (lower dot-dashed), $0.4$ (solid), $0.5$ (upper
dot-dashed), $\kappa=\kappa_0$, 
$M_s=0.4$ ,
$g(N)$ eq.~\gN\ with $r=1$
and
$\as=0.2$ compared with the NLO GLAP curve (dashed). The small kinks are due to
cuts at $N=\as c$ which reappear at this
accuracy through $\gamma_{ss}$. }}\hskip1truecm}
\endinsert
\topinsert
\vbox{
\epsfxsize=10truecm
\centerline{\epsfbox{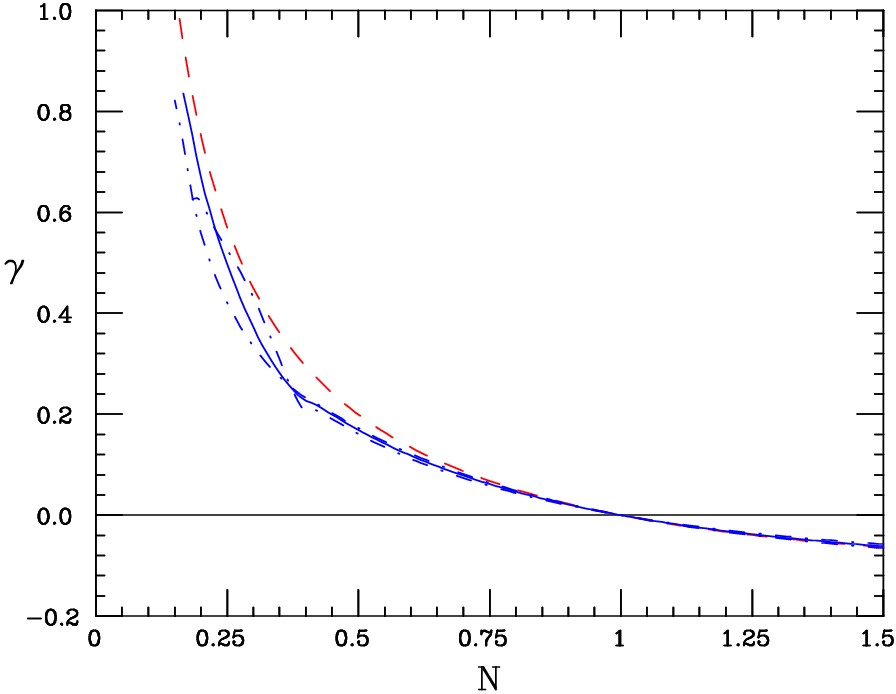}}
\hbox{
\vbox{\footnotefont\baselineskip6pt\narrower\noindent Figure 7: The improved
anomalous dimension eq.~\gamresexp\ for
$\as c=0.4$, $\kappa=5/4 \kappa_0$ (lower dot-dashed),
$\kappa=\kappa_0$ 
(solid), $\kappa=3/4 \kappa_0$
(upper dot-dashed),
$M_s=0.4$ ,
$g(N)$ eq.~\gN\ with $r=1$,
and
$\as=0.2$ compared with the NLO GLAP curve (dashed).
}}\hskip1truecm}
\endinsert
\topinsert
\vbox{
\epsfxsize=10truecm
\centerline{\epsfbox{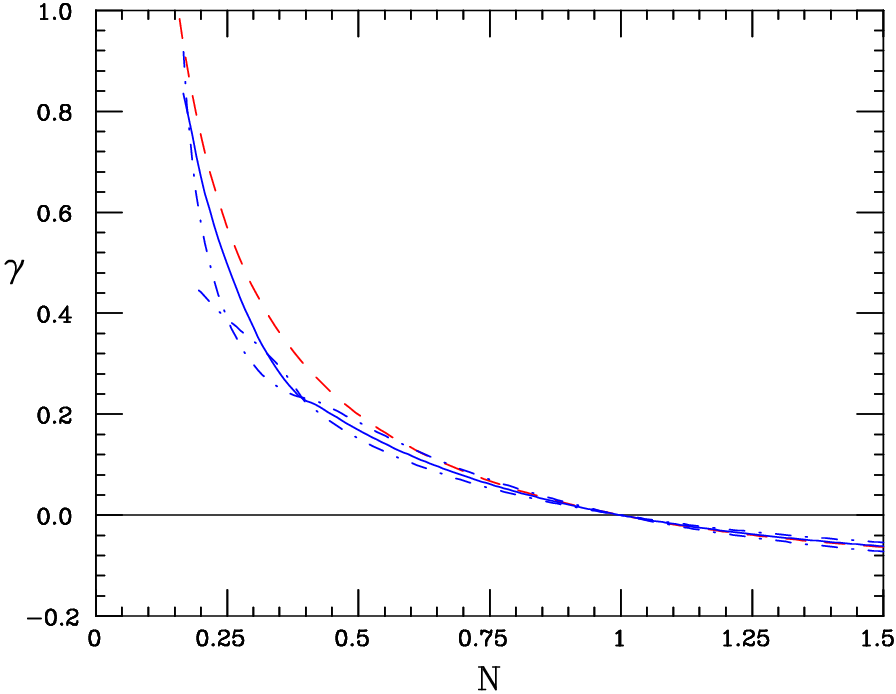}}
\hbox{
\vbox{\footnotefont\baselineskip6pt\narrower\noindent Figure 8: The improved
anomalous dimension eq.~\gamresexp\ for
$\as c=0.4$, $\kappa=\kappa_0$, $M_s=0.35$ (dot-dashed, lower at $N\sim0.5$), 
$M_s=0.4$ (solid),
$M_s=0.45$ (upper dot-dashed), 
$g(N)$ eq.~\gN\ with $r=1$,
and
$\as=0.2$ compared with the NLO GLAP curve (dashed).
}}\hskip1truecm}
\endinsert
The anomalous dimension eq.~\gamresexp\ is compared to its perturbative
counterpart in figs.~6-8, for typical values of
the parameters.
It is seen that a reasonable stability of the results is obtained for
$0.35\lsim M_s\lsim 0.45$, $0.3\lsim \as c\lsim 0.5$, $3/4 \kappa_0
\lsim \kappa \lsim 5/4 \kappa_0$, though at the extremes of the parameter range
some instability due to large subleading corrections starts to appear.
For instance, the discontinuity at $N=\as c$ 
in the first derivative of the running
coupling term eq.~\deltag\ starts showing up (figure~8) when $\kappa$
is lowered, since its coefficient is proportional to $1/\kappa$.

This shows that the leading order resummed expression
eq.~\leadimpr\ is stable upon perturbative
corrections for a range of values of the parameters which are reasonably close
to the leading order values
eq.~\mszero-\cappazero.  However, me must conclude that the subleading
correction eq.~\nlimprres\ cannot be used to
improve the leading order resummation, because the values of the parameters
which characterize it cannot be calculated in
a reliable way. In fact, as already mentioned, the NLO perturbative
$\chi=\chi_0+\as \chi_1$ has no minimum. As a consequence,
the uncertainty on this correction is of the same order as its size.

Hence, the available information on the parameters $c$, $\kappa$ and $M_s$
from the next-to-leading
order kernel $\chi_1$ is insufficient to improve the leading order
resummed result.
The only viable procedure, without further theoretical input, is
to combine the leading-order resummation eq.~\leadimpr\ with the
standard next-to-leading order correction to the GLAP equation, as we
did in eq.~\leadimprnl~(figure~5).

\newsec{Conclusion}

The main result of this work is the improved anomalous dimension given
in eq.~\leadimpr\ (and its generalizations eqs.
\leadimprdel\ and \leadimprnl). These relatively manageable expressions have
the correct perturbative limit for small
$\as$ with $N$ fixed (which corresponds to large $x$ for the splitting
functions) and include the leading BFKL
corrections with running coupling effects for small $\as$ with $\as/N$ fixed.
It is quite remarkable that the
improved anomalous dimension in eq.~\leadimpr, which does not
contain free parameters, leads to a splitting function which is surprisingly
close to the perturbative result down to small
values of
$x$ (figure 2) in agreement with the data. The running coupling
effects
are essential
to soften
the small-$x$ asymptotics, but {\it a priori} one would
not expect that the corrected
and the perturbative anomalous dimensions should
be so close over an extended range at small $x$. The typical BFKL rise is
softened and delayed to very small values of
$x$, while in the intermediate small-$x$ region the splitting function is flat
and close to the perturbative one. 

As is by now well
known, a flat perturbative splitting function reproduces by evolution a
rising structure function, and predicts
a double logarithmic slope in quantitative agreement with that seen in
the
data~\mf. However, the data suggest that the optimal value of the slope 
is contained in a region which is up to $15\%$ 
smaller than the leading order GLAP value. It follows that
the value of the splitting function in the flat
region, being proportional to the square of the slope, can be at most  
$30\%$ below its leading order GLAP value.
Thus it is quite non
trivial that our improved anomalous dimension directly fulfills this empirical
constraint in the flat region (where it
lies below the perturbative splitting function) down to
$x\sim 10^{-3}$. 

In comparison to the approach of refs.~\refs{\salam,\ciaf} we share the general
physical framework, but there are two main differences. The first
concerns the resummation procedure for $\chi(\as,M)$ near $M=0$. 
We match systematically the perturbative expansion of $\chi$ to that
of the usual anomalous dimension, thereby  finding a shift of order $\as$
of the poles of $\chi$ at
$M=0$ in a way which is made unambiguous by the
constraint of momentum conservation. They
instead, based on the change of scale variables in
going from the BFKL symmetric into the DIS
asymmetric configuration, displace the pole from
$M=0$ to
$M+N/2=0$. In the relevant limit,
$\as$ small with
$\as/N$ fixed, the two displacements are of the same order, but in
refs.~\refs{\salam,\ciaf}
the input of momentum conservation, which controls our double--leading
resummation, is not imposed
although a posteriori the results are numerically compatible with
it. Second, they
determine the asymptotic small-$x$ behaviour by
making some assumptions on the way to implement a symmetrisation $M \rightarrow
(1-M)$ of the BFKL kernel
$\chi(\alpha_s,M)$, which we prefer not to do because of the model
dependence it entails. For this we have to pay the price
of some parameter dependence in the non
leading terms. However, the fact that we find that after running coupling
resummation the parameter dependence is weak
gives support to their statement that the model dependence is not important.

However, an  advantage of our approach is
that we end up with an explicit formal expression for the improved anomalous
dimension which can be used in the
conventional evolution equations and thus matched to standard perturbative
evolution. In refs.~\refs{\salam,\ciaf} instead
only the resummed BFKL kernel is determined analytically, so anomalous
dimensions must be evaluated numerically from the
solution. This makes the possibility of fitting the data and matching to usual
evolution equations more difficult. 
The recent work~\newciaf\ by the same authors, which appeared
while the present work was being completed, adopts a somewhat different
symmetrisation procedure, such that, for example,
momentum conservation is not automatic and is
therefore imposed. 
A new result for the splitting function is found with a
dip in the intermediate region at
small $x$ before the onset of the BFKL rise at even smaller values of $x$. 
This confirms the model dependence of the approach of
refs.~\refs{\salam,\ciaf},  which
corresponds to the parameter dependence that we find. An
advantage of our approach in this respect, however,  
is that the parameter dependence can be
cleanly separated into the non-leading corrections.

Indeed, the simple leading order
expression of the improved anomalous dimension  eq. \leadimpr\ is only valid
when non leading corrections in both the
perturbative and the BFKL expansions can be neglected. We have discussed a
general procedure for a systematic treatment of
these higher order corrections, of which the NLO term is available, and we have
considered it in detail. Following a general
approach already used in our previous papers, our resummation strategy is to
reabsorb in the leading term the parameters that
describe the behaviour of $\chi(M)$ near $M=1/2$. In
refs.~\refs{\sxres,\sxphen}  we resummed the
value $\lambda=\as c$ determining the small-$x$
asymptotic behaviour as $x^{-\lambda}$. Here we include in the leading
$\chi(M)$ the first few terms of its expansion near the
minimum at
$M=M_s$ that are needed for the Airy resummation. From the known non-leading
correction we remove the perturbative expansion
of the parameters included in the leading term. While the non leading
correction is large before the shift we have shown that
it reduces to a small correction after the subtraction. Thus we have shown that
the non leading corrections are in principle
under control. 

The difficulty is that the negative infinite behaviour of
$\chi_1(M)$ near $M=1$ makes the treatment of the
non leading contribution in the central region of $M$ ambiguous. For
example, the perturbative curvature $\kappa$ becomes
negative in the central region when the $\chi_1$ term is added, while we assume
that the true curvature is positive
(corresponding to a minimum). On the basis of the closeness of the
leading--order result to standard perturbative evolution, and the
phenomenological success of the latter, 
we expect that the subleading correction
after subtraction is small. However, this also implies that it
contains ambiguities of the same order as the correction itself. 

Thus, in
practice,  the known next-to-leading order correction is mainly useful as an
estimate of the error on the leading approximation. The uncertainty on the non
leading correction can be phenomenologically
described in practice by replacing in the leading formula the lowest order
perturbative
values of the parameters describing the central region of $\chi$
with generic values to
be optimized to the data. For
example, we have explicitly examined the variation of our results
when the intercept $c$ of the anomalous dimension is
varied [eq.~\leadimprdel\ and figure 3], and we
also studied variations of the location of the minimum $M_s$ and the
curvature at the minimum $\kappa$.

In conclusion, we have achieved a precise quantitative understanding 
of the resummation of BFKL logarithms, which explains the empirical
observation of the validity of GLAP evolution  down to rather smaller
values of $x$ than might be naively expected. It will be interesting to
compare parton distribution functions evolved with our resummed 
anomalous dimension with the latest HERA data to look for 
evidence~\haidt\ for the small departures from GLAP evolution that we
predict.

\footatend\vfill\supereject\immediate\closeout\rfile\writestoppt
\baselineskip=14pt\centerline{{\bf References}}\bigskip{\frenchspacing%
\parindent=20pt\escapechar=` \input refs.tmp\vfill\eject}\nonfrenchspacing
\vfill\eject
\bye

\bigskip
\footatend\vfill\supereject\immediate\closeout\rfile\writestoppt
\baselineskip=14pt\centerline{{\bf References}}\bigskip{\frenchspacing%
\parindent=20pt\escapechar=` \input refs.tmp\vfill\eject}\nonfrenchspacing
\vfill\eject
\bye